\documentstyle[preprint,prb,aps]{revtex} 
\tighten 

\def\RR{\hbox{I \kern -.56em R}}                                
\def\CC{\hbox{\vrule height6.7pt width0.25pt \kern-.3em C}}     
\def\ZZ{\hbox{Z\kern-.33em Z}}                                  
\def\NN{\hbox{I \kern -0.55em N}}                               

\begin{document}
\draft

\title{Some Properties of Riesz Means
and Spectral Expansions}

\author{S. A. Fulling\\
{\it with an appendix by}\\    
R. A. Gustafson\\ \null{} }  

\address{Department of Mathematics\\
Texas A\&M University\\
College Station, Texas 77843-3368}

\date{October 6, 1997}

\maketitle

\begin{abstract}
  It is well known that short-time expansions of heat kernels 
correlate  to formal high-frequency expansions of spectral 
densities.  
 It is also well known that the latter expansions are generally not 
literally true beyond the first term.
 However, the terms in the heat-kernel expansion correspond 
rigorously  to quantities called Riesz means of the spectral 
expansion,
 which damp out oscillations in the spectral density at high 
frequencies by dint of performing an average over the density at 
all lower frequencies.
In general, a 
change of variables leads to new Riesz means that contain different 
information from the old ones.
 In particular, for the standard second-order elliptic operators, 
Riesz means with respect to the square root of the spectral 
parameter correspond to terms in the asymptotics of elliptic and 
hyperbolic Green functions associated with the operator, and these 
quantities contain ``nonlocal'' information not contained in the 
usual Riesz means and their correlates in the heat kernel.
 Here the relationship between these two sets of Riesz means is 
worked out in detail;
 this involves just classical one-dimensional analysis and 
calculation, with no substantive input from spectral theory or 
quantum field theory.
 This work provides a general framework for calculations that are 
often carried out piecemeal (and without precise understanding of 
their rigorous meaning) in the physics literature.
\end{abstract}


\vfill
E-mail: {\tt fulling@math.tamu.edu} 

\section{Physical motivation}\label{sec1}

Let $H$ be a positive, self-adjoint, elliptic, second-order partial 
differential operator. 
 For temporary expository simplicity, assume that $H$ has purely 
discrete spectrum, with eigenvalues $\lambda_n$ and normalized 
eigenfunctions $\phi_n(x)$. 
 Quantum field theorists, especially those working in curved space, 
are accustomed to calculating (1) the {\em heat kernel}, 
\begin{equation} 
 K(t,x,y) = \sum^\infty_{n=1} \phi_n(x) \phi_n(y)^* e ^{-t\lambda_n};
\label{(1.1)}\end{equation}
(2) the {\em Wightman function},
\begin{equation}
W(t,x,y) = \sum^\infty_{n=1} \phi_n(x) \phi_n(y)^* {1\over
2\sqrt{\lambda_n}} e^{-it\sqrt{\lambda_n}},
 \label{(1.2)}\end{equation}
which determines the vacuum energy density of a quantized field 
 in a static space-time background. 
 Of greatest interest are the behaviors of these
functions with $y=x$ 
 (either evaluated pointwise, or integrated over the
whole spatial manifold) in the limit $t\to 0$.

Roughly speaking, the small-$t$ asymptotics of both
  (\ref{(1.1)}) and (\ref{(1.2)}) are
determined by the large-$\lambda$ (ultraviolet) asymptotics of the 
density of eigenvalues and of the eigenfunctions. 
 However, there is a major difference between the heat kernel and 
the Wightman function in this regard. 
  {\em All\/} the coefficients $b_s(x)$ in the heat-kernel expansion
\begin{equation}
 K(t,x,x) \sim \sum^\infty_{s=0} b_s(x) t^{-{m\over 2}+{s\over 2}}
 \label{(1.3)}\end{equation}
($m=$ dimension) are {\em locally determined\/} by the coefficient 
functions in the differential operator $H$ {\em at the point\/} 
$x$. 
 The same is true of the leading, singular terms in an expansion of 
$W(t,x,y)$, which are removed by renormalization; 
 the finite residue, however, is a {\em nonlocal\/} functional of 
the coefficients in $H$, containing information about, for example, 
the global structure of the manifold --- this is what makes vacuum 
energy interesting and nontrivial. 
 It follows that this nonlocal information is somehow contained in 
the ultraviolet asymptotics of the spectrum, although it is lost in 
passing (pointwise) to $K(t,x,x)$. 
  The Wightman function is more typical of Green functions 
  associated with $H$; 
 the striking and somewhat mysterious thing is how {\em special\/} 
the heat kernel is. 

{\em The primary purpose of this paper is to point out that many of 
the facts of this subject have nothing to do specifically with 
partial differential operators (much less with quantum field 
theory). 
 Rather, they result from some classical real analysis (in one 
dimension) concerning the summability of series and integrals, much 
of which was developed by M.~Riesz and G.~H.~Hardy early in the 
twentieth century\/}. 
 One consequence is that many formal relationships between the 
asymptotic expansions of Green functions and those of the 
associated spectral measures, and between the asymptotic expansion 
of one Green function and that of another, can be worked out in the 
abstract without reference to the detailed spectral theory of any 
particular operator. 
 The same is true, qualitatively, of the {\em limitations\/} of such 
relationships:  
 The rigorous asymptotic expansion of a heat kernel suggests a 
formal asymptotic expansion of an associated spectral measure, but 
the latter is usually valid only in some averaged sense. 
 It can be translated into rigorous statements about the {\em Riesz 
means\/} of the measure. 
 The construction of Riesz means not only washes out oscillations in 
the measure at the ultraviolet end, but also incorporates some 
information about the measure at small or moderate values of 
$\lambda$. 
 The information contained in a Riesz mean depends on the variable 
of integration; 
  for example, the Riesz means with respect to 
$\sqrt\lambda$ contain more information than those with respect to 
$\lambda$. 
 The difference between the nonlocal asymptotics of the Wightman 
function and the local asymptotics of the heat kernel is, at root, 
an example of this phenomenon. 

These relationships can be made yet more precise by introducing 
some concepts from distribution theory\cite{EF,EGV}. 

\section{Survey of the phenomena}\label{sec2}

To provide a concrete context for the later sections of the paper, 
we display here some Green functions and asymptotic expansions 
associated with a constant-coefficient differential operator in one 
dimension. 
 (The later sections are logically independent of this 
one and much more general). 

Here and in Sections \ref{sec6} and \ref{sec7}
we use as a surrogate for the Wightman function a
technically simpler Green function that we call the 
 {\em cylinder kernel}. 
 This is the integral kernel $T(t,x,y)$ of the operator
$e^{-t\sqrt H}$, and it is related to the elliptic equation
\begin{equation}
 {\partial^2\Psi\over \partial t^2} = H\Psi
 \label{(2.1)}\end{equation}
in the same way that the heat kernel
  (the integral kernel of $e^{-tH}$) 
 is related to the parabolic equation
 \begin{equation} 
 -\,{\partial\Psi\over \partial t} = H\Psi.
 \label{(2.2)}\end{equation}
That is,
\begin{equation}
 \Psi(t,x) = \int_{\cal M} T(t,x,y) f(y)\,dy
 \label{(2.3)}\end{equation}
solves (\ref{(2.1)}) in the infinite half-cylinder 
 $(0,\infty) \times {\cal M}$
with the initial data
\begin{equation}
 \lim_{t\downarrow 0} \Psi(t,x) = f(x)
 \label{(2.4)}\end{equation}
on the manifold ${\cal M}$. 
 The cylinder kernel displays the same kind of
nonlocal short-time asymptotics as the Wightman function.

In this section we consider
\begin{equation}
 H = -\,{\partial^2\over \partial x^2}
 \label{(2.5)}\end{equation}
on various one-dimensional manifolds ${\cal M}$.

\subsection*{Case ${\cal M}=\RR$}

\nobreak
The heat kernel is
\begin{equation}
 K(t,x,y) = (4\pi t)^{-1/2} e^{-(x-y)^2/4t}.
 \label{(2.6)}\end{equation}
As $t\downarrow 0$,
\begin{equation}
 K(t,x,y) \sim 
 \cases{
 (4\pi t)^{-1/2} + O(t^\infty)&\text{if $y=x$,}\cr
\noalign{\smallskip}
O(t^\infty)&\text{if $y\ne x$;} }
 \label{(2.7)}\end{equation}
that is, all terms in the expansion (\ref{(1.3)}) beyond the first vanish.

The cylinder kernel is
\begin{equation}
 T(t,x,y) = {t\over\pi} {1\over(x-y)^2 +t^2}\,.
 \label{(2.8)}\end{equation}
As $t\downarrow 0$,
\begin{equation}
 T(t,x,y) \sim 
\cases{
 \displaystyle {1\over \pi t}&\text{if $y=x$,}\cr
\noalign{\smallskip}
 \displaystyle
{t\over \pi(x-y)^2}\sum\limits^\infty_{k=0} (-1)^k \left({t\over
x-y}\right)^{2k}& \text{if $y\ne x$.} }
 \label{(2.9)}\end{equation}
 (For the distributional, rather than pointwise, limit, 
 see Ref.~\onlinecite{EF}.)

\subsection*{Case ${\cal M} = \RR^+$}

\nobreak
We consider the operator (\ref{(2.5)}) on the interval $(0,\infty)$, 
with either the Dirichlet or the Neumann boundary condition at 0. 
  The Green functions 
are most easily obtained by the method of images from the previous 
case. 
 The heat kernel is
\begin{equation}
 K(t,x,y) = (4\pi t)^{-1/2} \left[e^{-(x-y)^2/4t} \mp
e^{-(x+y)^2/4t}\right],
  \label{(2.10)}\end{equation}
where {\em the upper and lower signs are for the Dirichlet and Neumann
cases, respectively}. 
 Because of the rapid decay of the image term, the
asymptotic behavior is still described exactly by (\ref{(2.7)}); 
 the heat kernel
in the interior does not sense the existence of the boundary. 
 The cylinder
kernel is
\begin{equation}
 T(t,x,y) = {t\over\pi} \left[{1\over (x-y)^2+t^2} \mp {1\over (x+y)^2 +
t^2}\right].
 \label{(2.11)}\end{equation}
As $t\downarrow 0$,
\begin{equation}
 T(t,x,y) \sim 
 \cases{
 \displaystyle{1\over\pi t}\mp {t\over \pi(2x)^2}
\sum\limits^\infty_{k=0} (-1)^k \left({t\over 2x}\right)^{2k}
 &\text{if $y=x$,}\cr
\noalign{\smallskip}
 \displaystyle
{t\over \pi(x-y)^2} \sum\limits^\infty_{k=0} (-1)^k \left({t\over
x-y}\right)^{2k}
\mp {t\over \pi(x+y)} \sum\limits^\infty_{k=0} (-1)^k \left({t\over
x+y}\right)^{2k}&\text{if $y\ne x$.} }
 \label{(2.12)}\end{equation}
Because of the slow decay of the basic kernel in its role as image 
term, the expansion differs from (\ref{(2.9)}) beyond the 
 leading $O(t^{-1})$ term; 
 the cylinder kernel senses the presence of the boundary, the type 
of boundary condition, and the distance to the boundary 
 (more precisely, the length, $x+y$, of a path between the two 
arguments that bounces off the endpoint\cite{F3}).

\subsection*{Case ${\cal M}=S^1$}

\nobreak
Consider (\ref{(2.5)}) on the interval $(-L,L)$ 
 with periodic boundary conditions.
The heat kernel is the well known theta function\cite{DMcK}
\begin{equation}
 K(t,x,y) = (4\pi t)^{-1/2} \sum^\infty_{n=-\infty} e^{-(x-y-2nL)^2/4t}.
\label{(2.13)}\end{equation}
The expansion (\ref{(2.7)}) is still valid.
  The cylinder kernel could also be
expressed as an infinite image sum, 
 but its Fourier representation can be
expressed in closed form (via the geometric series):
\begin{eqnarray}
T(t,x,y) &=& {1\over 2L} \sum^\infty_{k=-\infty} e^{i\pi
k(x-y)/L} e^{-\pi |k|t/L}\nonumber\\*
&=& {1\over 2L} {\sinh(\pi t/L)\over \cosh(\pi t/L) - \cos
(\pi(x-y)/L)}\,.
 \label{(2.14)}\end{eqnarray}
The first few terms of the expansion as $t\downarrow 0$ are
\begin{equation}
 T(t,x,y) \sim 
 \cases{
 \displaystyle{1\over\pi t} \left[1 +{1\over 12}
\left({\pi t\over
L}\right)^2 - {1\over 720} \left({\pi t\over L}\right)^4 
 + O(t^6)\right]&
\text{if $y=x$,}\cr 
\noalign{\smallskip}
 \displaystyle
{\pi t\over 2L^2} {1\over 1-\kappa}
  \left[1 + \left({\pi t\over L}\right)^2
\left({1\over 6} - {1\over 2(1-\kappa)}\right) 
 + O(t^4)\right]&\text{if $y\ne x$,} }
  \label{(2.15)}\end{equation}
where
\begin{equation}
 \kappa = \cos {\pi(x-y)\over L}\,.
 \label{(2.16)}\end{equation}
Thus the cylinder kernel is locally sensitive to the size of the 
manifold. 
 (In the limit of large $L$, (\ref{(2.15)}) matches (\ref{(2.9)}),
  as it should.)

In this case it is possible to ``trace'' the Green functions over the
manifold:
\begin{equation}
 K(t) \equiv \int^L_{-L} K(t,x,x)\,dx
\sim (4\pi t)^{-1/2} (2L) + O(t^\infty)
 \label{(2.17)}\end{equation}
(since (\ref{(2.7)}) is uniform in ${\cal M}$), and
\begin{equation}
T(t) \equiv \int^L_{-L} T(t,x,x)\,dx = {\sinh (\pi t/L)\over
\cosh(\pi t/L) -1}
\sim {2L\over \pi t} \left[ 1 + {1\over 12} \left({\pi t\over L}\right)^2
+\cdots \right].
 \label{(2.18)}\end{equation}
Both have leading terms proportional to the volume of the manifold, but
(\ref{(2.18)}) has higher-order correction terms analogous to those in 
 (\ref{(2.9)}).

\subsection*{Case ${\cal M}=I$}

\nobreak
We consider (\ref{(2.5)}) on the interval $I = (0,L)$. 
 For brevity we consider only
Dirichlet boundary conditions and the expansions on the diagonal
(coincidence limit).
 The heat kernel is
\begin{equation}
 K(t,x,y) = (4\pi t)^{-1/2} \sum^\infty_{n=-\infty}
\left[e^{-(x-y-2nL)^2/4t} - e^{-(x+y-2nL)^2/4t}\right].
 \label{(2.19)}\end{equation}
Expansion (\ref{(2.7)}) is valid in the interior 
 (but not uniformly near the endpoints). 
 The cylinder kernel is 
\begin{equation} 
 T(t,x,y) = {1\over 2L} \left[{\sinh(\pi t/L)\over \cosh(\pi t/L) -
\cos(\pi(x-y)/L)} - {\sinh(\pi t/L)\over \cosh(\pi t/L) - \cos(\pi(x+y)/L)}
\right].
 \label{(2.20)}\end{equation}
Not surprisingly, its expansion combines the features of 
 (\ref{(2.12)}) and (\ref{(2.15)}).
\begin{eqnarray}
T(t,x,x) &\sim& {1\over \pi t} \left[1 +{1\over 12} \left({\pi
t\over L}\right)^2 - {1\over 720} \left({\pi t\over L}\right)^4
+\cdots\right]
 \nonumber\\*
&&- {\pi t\over 2L^2} {1\over 1-\cos( 2\pi x/L)} \left[1 +
\left({\pi t\over L}\right)^2 \left({1\over 6} - {1\over 2(1-\cos(2\pi
x/L))}\right)+\cdots\right]
 \nonumber\\
&=& {1\over \pi t} \left[1 + \left({\pi t\over L}\right)^2 \left({1\over 12}
- {1\over 2(1-\cos(2\pi x/L))}\right)\right.\nonumber\\*
&& \left.+ \left({\pi t\over L}\right)^4 \left({-1\over 720} - {1\over
12(1-\cos(2\pi x/L))} + {1\over 4(1-\cos(2\pi x/L))^2}\right) +\cdots
\right].
 \label{(2.21)}\end{eqnarray}
(Compare (\ref{(2.12)}) in the form
\begin{equation}
 T(t,x,x) \sim {1\over \pi t} \left[1 - {t^2\over 4x^2} + {t^4\over 16
x^4} +\cdots\right],
 \label{(2.22)}\end{equation}
which (\ref{(2.21)}) matches as $x\to 0$ or $L\to \infty$.)

For the traces in this case one has
\begin{equation}
K(t) \equiv  \int_0^L K(t,x,x)\, dx 
\sim (4\pi t)^{-1/2} L -{\textstyle {1\over2}} + O(t^\infty),
 \label{(2.23)} \end{equation}
a well known result\cite{G79}, 
 and
\begin{eqnarray}
 T(t) &\equiv& \int_0^L T(t,x,x)\, dx 
 \nonumber\\*
&=& {1\over2}\, {\sinh(\pi t/L) \over \cosh(\pi t/L) -1 }
- {\sinh(\pi t/L) \over2L} \int_0^L 
 {dx \over\cosh(\pi t/L) - \cos(2\pi x/L)}
\nonumber\\
&=& {1\over2}\, {\sinh(\pi t/L) \over \cosh(\pi t/L) -1 } -{1\over2} 
 \nonumber\\*
&\sim& {L\over \pi t} \left[ 1 - {\pi t \over 2L} - {1\over12}
\left( {\pi t\over L}\right)^2 +O(t^4) \right]. 
 \label{(2.24)}\end{eqnarray}
In comparison with (\ref{(2.17)}) and (\ref{(2.18)}),
   the leading terms of (\ref{(2.23)}) and
(\ref{(2.24)}) have adjusted to reflect the smaller size of the 
manifold, 
 and the new second ($t$-independent) terms are the effect 
of the boundary. 
 The minus sign on those terms distinguishes the 
Dirichlet boundary condition from the Neumann.  
 (A less trivial 
differential operator would yield more complicated expansions, 
 with higher-order terms exhibiting an interaction between the 
boundary condition and the coefficients (potential, curvature) 
 (see Ref.~\onlinecite{G79}).) 

It is worth noting that the most important dependence of 
(\ref{(2.18)}) or (\ref{(2.24)}) on the size of the manifold comes from 
the integration, not from the form of the integrand. 
 (Indeed, the $L$-dependence of (\ref{(2.15)}) and (\ref{(2.21)}) as 
written is downright misleading in this respect.) 
 In ignorance of 
both the size of ${\cal M}$ and the nature and location of any 
boundaries, knowledge of the $O(t)$ term in $T(t,x,x)$ on a small 
interval of~$x$ would be a rather useless tool for the inverse 
problem. 

\section{Notation and basic formulas} \label{sec3}

The basic references for this section and much of the next are 
 Hardy\cite{Ha1,HR,Ha2} and H\"or\-man\-der\cite{Ho1,Ho2}.
 (The formulations given here are somewhat new, however.)

Let $f(\lambda)$ be a function of locally bounded variation on 
 $[0,\infty)$
such that
\begin{equation}
 f(0) = 0.
 \label{(3.1)}\end{equation}
Typically, $f$ will be defined as a Stieltjes integral
\begin{equation}
 f(\lambda) = \int^\lambda_0 a(\sigma)\, d\mu(\sigma),
 \label{(3.2)}\end{equation}
where $\mu(\lambda)$ is another function of the same kind, 
 and $a(\lambda)$ is (say) a continuous function. 
 By convention we take functions of locally
bounded variation to be continuous from the left:
\begin{equation}
 f(\lambda) \equiv \lim_{\varepsilon\downarrow 0}
f(\lambda-\varepsilon);  
 \label{(3.3)}\end{equation}
 \begin{equation}
\int^b_a a(\sigma)\, d\mu(\sigma) \equiv \lim_{\varepsilon\downarrow 0}
\int^{b-\varepsilon}_{a-\varepsilon} a(\sigma)\,d\mu(\sigma).
 \label{(3.4)}\end{equation}

By $\partial^\alpha_\lambda f$ we denote the $\alpha$th 
 derivative of $f$ with respect to $\lambda$. 
 Since iterated indefinite integrals can be
regarded as derivatives of negative order, we define
\begin{equation}
 \partial^{-\alpha}_\lambda f(\lambda) 
 \equiv \int^\lambda_0 d\sigma_1
\cdots \int^{\sigma_{\alpha-1}}_0 d\sigma_\alpha 
 \,f(\sigma_\alpha). 
 \label{(3.5)}\end{equation}
As is well known, the iterated integral is equal to the single integral
\begin{equation}
 \partial^{-\alpha}_\lambda f(\lambda) = {1\over (\alpha-1)!}
\int^\lambda_0 (\lambda-\sigma)^{\alpha-1} f(\sigma)\,d\sigma,
 \label{(3.6)}\end{equation}
which in turn may be converted to
\begin{equation}
 \partial^{-\alpha}_\lambda f(\lambda) = {1\over \alpha!} \int^\lambda_0
(\lambda-\sigma)^\alpha \,d f(\sigma).
 \label{(3.7)}\end{equation}
This last formula remains meaningful for $\alpha=0$, yielding the 
natural definition 
 \begin{equation}
 \partial^0_\lambda f(\lambda) = \int^\lambda_0 df(\sigma) = f(\lambda).
\label{(3.8)}\end{equation}

The $\alpha$th {\em Riesz mean\/} of $f$ is defined by
\begin{equation}
 R^\alpha_\lambda f(\lambda) \equiv \alpha! \,\lambda^{-\alpha}
\partial^{-\alpha}_\lambda f(\lambda) = \int^\lambda_0 \left(1 -
{\sigma\over \lambda}\right) ^\alpha df(\sigma).
 \label{(3.9)}\end{equation}
We call $\partial^{-\alpha}_\lambda f(\lambda)$ the $\alpha$th 
 {\em Riesz integral\/} of $f$, and we call
\begin{equation}
 \lim_{\lambda\to \infty} R^\alpha_\lambda f(\lambda),
 \label{(3.10)}\end{equation}
if it exists, the $\alpha$th {\em Riesz limit\/} of $f$. 
 If such a limit
exists for $f$ defined by (\ref{(3.2)}),
  we say that the integral (\ref{(3.2)}) is
  {\em summable by  Riesz means of order\/} $\alpha$. 
 Eq.~(\ref{(3.7)}) may be used to
define all these things for any 
 $\alpha \in \CC$ with $\mathop{\rm Re} \alpha >-1$, 
 but {\em here we shall consider only nonnegative integer 
$\alpha$.}

 From (\ref{(3.5)}) and (\ref{(3.6)})
  with $f$ replaced by $\partial^{-\beta}_\lambda f$, we
have (for $\alpha>0$, $\beta\ge 0$)
\begin{equation}
 \partial^{-(\alpha+\beta)}_\lambda f(\lambda) = {1\over (\alpha-1)!}
\int^\lambda_0 (\lambda-\sigma)^{\alpha-1} \partial^{-\beta}_\sigma
f(\sigma)\,d\sigma,
 \label{(3.11)}\end{equation}
hence
\begin{equation}
 R^{\alpha+\beta}_\lambda f(\lambda) = \alpha {\alpha+\beta \choose
\alpha}
\lambda^{-(\alpha+\beta)} \int^\lambda_0 (\lambda-\sigma)^{\alpha-1}
\sigma^\beta R^\beta_\sigma f(\sigma)\,d\sigma.
 \label{(3.12)}\end{equation}
 From this is proved Hardy's {\em first theorem of consistency\/}:
\begin{equation}
 \qquad\quad\vtop{\advance\hsize -1.5truein \noindent
If an integral (\ref{(3.2)}) is summable by Riesz
means of order $\alpha_0\,$, then it is summable
to the same value of means of order $ \alpha$
for any $\alpha\ge \alpha_0\,$. 
 In particular, if the
integral is convergent, then it is summable
by means of any order  $\alpha\ge 0$.}
 \label{(3.13)}\end{equation}
The proof is  similar to that of (\ref{(4.5)}), below.

 Although Riesz and Hardy were primarily interested in defining 
numerical values for nonconvergent series and integrals,
 our concern (following H\"or\-man\-der) is to use Riesz means as a way 
of organizing the asymptotic information contained in a function, 
without interference from small-scale fluctuations.
 Thus we are more interested in the Riesz means themselves than in 
the Riesz limits.

\section{The relation between means with respect to different 
variables} \label{sec4}

Let $\omega$ be related to $\lambda$ by
\begin{equation}
 \lambda  = \omega^k \qquad (k>0,\ k\ne 1).
 \label{(4.1)}\end{equation}
We are primarily interested in the cases $k=2$ and $k = {1\over 2}$.
  (More generally, one could treat two variables related by any
orientation-preserving diffeomorphism of $[0,\infty)$.) 
 We write $\sigma$
and $\tau$ as integration variables corresponding to 
 $\lambda$ and $\omega$
respectively --- hence $\sigma = \tau^k$. 
 We let
\begin{equation}
 \tilde f(\omega) \equiv f(\omega^k) = f(\lambda)
 \label{(4.2)}\end{equation}
and may omit tildes when no confusion seems likely.

The Riesz mean $R^\alpha_\lambda f$ can be expressed in terms of 
the mean $R^\alpha_\omega f$ [${}\equiv R^\alpha_\omega \tilde f$] 
and vice versa. 
 We
call the following the {\em Hardy formula\/} because it is implicit 
in  Hardy's 1916 proof of the ``second theorem of consistency'':
\begin{equation}
 R^\alpha_\lambda f = k^\alpha R^\alpha_\omega f + \int^\lambda_0
J_{k,\alpha}(\lambda,\sigma) R^\alpha_\tau f\, d\sigma,
 \label{(4.3)}\end{equation}
where
\begin{equation}
 J_{k,\alpha}(\lambda,\sigma) \equiv \sum^{\alpha-1}_{j=0}
  (-1)^{\alpha-j} \lambda^{-j-1} {1\over j!\,(\alpha-1-j)!} 
 {\Gamma(kj+k)\over \Gamma(kj +k-\alpha)}\sigma^j.
 \label{(4.4)}\end{equation}
In (\ref{(4.3)}), $R^\alpha_\tau f$ means $R^\alpha_\tau\tilde f$ 
 evaluated at $\tau= \sigma^{1/k}$. 
 In (\ref{(4.4)}), the ratio of $\Gamma$ functions is interpreted
as $0$ if $kj +k-\alpha$ is a nonpositive integer.

{\em Proof:}
  Use successively the definitions 
 (\ref{(3.9)}), (\ref{(3.6)}), (\ref{(4.1)}), and (\ref{(3.5)}):
\begin{eqnarray*}
R^\alpha_\lambda f &=& \alpha! \lambda^{-\alpha}
\partial^{-\alpha}_\lambda f\\*
&=& \alpha \lambda^{-\alpha} \int^\lambda_0 (\lambda-\sigma)^{\alpha-1}
f(\sigma)\,d\sigma\\
&=& \alpha \omega^{-\alpha k} \int^\omega_0 (\omega^k-\tau^k)^{\alpha-1}
k\tau^{k-1} \tilde f(\tau) \,d\tau\\*
&=& \alpha \omega^{-\alpha k} \int^\omega_0 (\omega^k-\tau^k)^{\alpha-1}
k\tau^{k-1} \partial^\alpha_\tau(\partial^{-\alpha}_\tau\tilde f)\,d\tau.
 \end{eqnarray*}
Integrate by parts $\alpha$ times. 
 The lower-endpoint term always vanishes,
because $\partial^{-\beta}_\omega f(0) = 0$ for all $\beta$. 
 Until the last step, the upper-endpoint term contains positive 
powers of 
 $\omega^k-\tau^k$, so it also vanishes. 
 At the last step there is an upper-endpoint contribution 
 \[ 
R_1 = (-1)^{\alpha-1} \alpha \omega^{-\alpha k}(\alpha-1)!\,
(-k\tau^{k-1})^\alpha (-1) \partial^{-\alpha}_\tau \tilde
f\Big|_{\tau=\omega}
= \alpha!\,k^\alpha \omega^{-\alpha} \partial^{-\alpha}_\omega f
  = k^\alpha R^\alpha_\omega f. 
\]
The remaining integral is
\[ R_2 = (-1)^\alpha \alpha \omega^{-\alpha k} \int^\omega_0
\partial^\alpha_\tau
[(\omega^k-\tau^k)^{\alpha-1}k\tau^{k-1}]
 \partial^{-\alpha}_\tau \tilde f\, d\tau.
 \]
But
\[ (\omega^k-\tau^k)^{\alpha-1} k\tau^{k-1} 
 = k \sum^{\alpha-1}_{j=0} (-1)^j
{\alpha-1\choose j} \omega^{k(\alpha-1-j)} \tau^{kj+k-1},
\]
so
\begin{eqnarray*}
&&(-1)^\alpha \alpha\omega^{-\alpha k} \partial^\alpha_\tau
[(\omega^k-\tau^k)^{\alpha-1}k\tau^{k-1}]\\*
&=&\alpha k \sum^{\alpha-1}_{j=0} (-1)^{\alpha-j}
  {\alpha-1\choose j}
\left[\prod^{\alpha-1}_{i=0} (kj+k-1-i)\right] \tau^{kj+k-1-\alpha}
\omega^{-kj-k}.
 \end{eqnarray*}
Thus, by (\ref{(3.9)}) applied to $\tilde f(\tau)$,
\begin{eqnarray*}
R_2 &=& \int^\omega_0 \sum^{\alpha-1}_{j=0} {(-1)^{\alpha-j}
\alpha! \over j!\,(\alpha-1-j)!} 
 {\Gamma(kj+k)\over \Gamma(kj+k-\alpha)}\,
\tau^{kj} \partial_\tau(\tau^k) \omega^{-kj-k} {1\over \alpha!}
R^\alpha_\tau \tilde f\, d\tau\\*
&=& \int^\lambda_0 J_{k,\alpha}(\lambda,\sigma) R^\alpha_\tau f\,
d\sigma.
 \end{eqnarray*}
Adding $R_1$ and $R_2\,$, one obtains the formula (\ref{(4.3)})
  to be proved.

As a corollary we see:
\begin{equation}
 \qquad\quad\vtop{\advance\hsize -1.5truein \noindent
If an integral (\ref{(3.2)}) is summable by
$\alpha$th-order
Riesz means with respect to $\lambda$, then it is
summable to the same value by $\alpha$th means
with  respect to $\omega$, and conversely.} 
 \label{(4.5)}\end{equation}

{\em Proof:\/} We show the converse;
 the direct statement  follows when one replaces $k$ by $1/k$ 
and interchanges $\lambda$ and $\omega$. 
It suffices, since Riesz means are linear, 
 to consider the special cases 
 (i) where the Riesz limit in question is 0, and 
 (ii) where $f(\lambda) = C$, a
constant, for all $\lambda\ne 0$. 
 The latter case is trivial since all
Riesz means and hence all Riesz limits are equal to~$C$. 
 In case (i), we are given that
 $R^\alpha_\tau f \to 0$ as $\sigma\to \infty$; 
 that is, for every
$\varepsilon>0$ there is a $K$ such that 
 $|R^\alpha_\tau f(\sigma)|<\varepsilon$ 
 when $\sigma>K$. 
 For $\lambda>K$, write
\[
 R_2 = \int^K_0 J_{k,\alpha}(\lambda,\sigma) R^\alpha_\tau f\,
d\sigma + \int^\lambda_K J_{k,\alpha} (\lambda,\sigma) R^\alpha_\tau f\,
d\sigma
\equiv R_{21} + R_{22}\,.
 \]
 From the form of $J_{k,\alpha}$ (\ref{(4.4)}),
  $R_{21}$ approaches 0 as $\lambda\to
\infty$, and $|R_{22}|$ has the form
\[ |R_{22}| \le 
 \varepsilon \sum^{\alpha-1}_{j=0} c_j \lambda^{-j-1} \int^\lambda_K
\sigma^j\, d\sigma = \varepsilon O(\lambda^0),
\] 
which can be made arbitrarily small.
 Thus $R^\alpha_\lambda f$ approaches~0, as asserted.


A generalization of this argument 
 (due to Riesz and  Hardy\cite{Ha1,HR}) proves the
  {\em second theorem of consistency\/} (for $\alpha$ integer):
Let $\lambda=g(\omega)$, where $g$ is an
increasing
function in $C^\infty(\RR^+)$, $g(0) = 0$, $g(\infty) = \infty$, and
\begin{equation}
 \partial^r_\omega g(\omega) = O(\omega^{-r}g(\omega)) 
\quad\text{for all $r=1,2,\ldots.$}
 \label{(4.7)}\end{equation}
Then (\ref{(4.5)}) holds, 
 except possibly for the clause ``and conversely''\negthinspace.
Conditions sufficient together to guarantee (\ref{(4.7)}) 
 (given the other conditions on $g$) are
\begin{equation}
 g(\omega) = O(\omega^\Delta) \quad \text{for some $\Delta>0$} 
 \label{(4.8)}\end{equation}
and
\begin{equation}
 \qquad\quad\vtop{\advance\hsize -1.5truein \noindent
$g(\omega)$  is given (for sufficiently large
$\omega$) by an
explicit, finite formula involving the logarithmic
and exponential functions, real constants,
and (real, finite) elementary algebraic operations.}
 \label{(4.9)}\end{equation}

The significance of this theorem is clearer in the following 
imprecise paraphrase: 
  (1)~If $\omega$ increases to $\infty$, but {\em more
slowly\/} than $\lambda$, 
 then $\lambda$-summability implies $\omega$-summability,
  {\em provided\/} that the increase of $\omega$ with
$\lambda$ is sufficiently ``steady'' --- 
 that is, derivatives of $g$ must not
oscillate so as to disrupt (\ref{(4.7)}).
  (2)~Under those conditions,
$\omega$-summability does not ensure $\lambda$-summability --- 
 e.g., an integral $R^\alpha$-summable with respect to $\omega$ may 
not be $R^\alpha$-summable with respect to $\lambda\sim e^\omega$. 
 However, if the rates of increase of $\lambda$ and $\omega$ differ 
only by a power, then the two types of summability are coextensive. 

If $k$ is an integer (necessarily $\ge 2$), there is another 
relation between $\lambda$-means and $\omega$-means, which we call 
the {\em H\"or\-man\-der formula\/}
 [Ref.~\onlinecite{Ho2}, Sec.~5]: 
 \begin{eqnarray}
R^\alpha_\lambda f(\lambda) &=& \int^\omega_0\left[\sum^k_{j=1}
(-1)^{j-1} {k\choose j} 
 \left(1-{\tau\over\omega}\right)^j\right]^\alpha
d\tilde f(\tau) \nonumber\\*
&=& \sum^{\alpha k}_{\beta=\alpha} b_\beta  R^\beta _\omega f\quad 
 \text{for certain numbers }
b_\beta \,. \label{(4.10)}
 \end{eqnarray}
This is obtained from (\ref{(3.9)}) by expanding the factor 
 $(1-\sigma/\lambda) = 1 - [1-(1-\tau/\omega)]^k$ 
 by the binomial theorem. 
 In view of (\ref{(3.13)}), (\ref{(4.10)})
has the converse part of (\ref{(4.5)}) as a corollary.
  More significantly, the
absence of an integral in (\ref{(4.10)}) as compared with 
 (\ref{(4.3)}) means that the asymptotic $(\lambda\to \infty)$ 
behavior of $R^\alpha_\lambda f$ is entirely determined by the {\em 
asymptotic\/} behavior of $R^\beta_\omega f$, 
 a conclusion which we shall reach independently below. 
 We shall see also that the converse is false:  The asymptotic 
behavior of $R^\alpha_\omega f$ is affected by the values of 
$f$ at {\em small\/} $\lambda$,
 in a way that is not captured by the asymptotic behavior of 
$R^\beta_\lambda f$ for any $\beta$, no matter how large. 
 Thus it is really essential in (\ref{(4.10)}) that $k$ is
an integer (so that the binomial series terminates).

 We now work out in each direction the relation between the Riesz 
means with respect to the spectral parameter and those with respect 
to its square root, taking note of a fundamental difference between 
the two calculations.

  \paragraph*{\bf 1.}
 First assume that as $\lambda\to \infty$,
\begin{equation}
 R^\alpha_\lambda\mu(\lambda) 
 = \sum^\alpha_{s=0} a_{\alpha s} \lambda^{{m \over 2}-{s\over 2}} 
 + O(\lambda^{{m\over 2}-{\alpha\over 2}-{1\over 2}})
\label{(4.11)}\end{equation}
for some positive integer $m$, 
 as suggested by H\"or\-man\-der's results on the
Riesz means of spectral functions of second-order operators 
 (where $m$ is the dimension of the underlying manifold). 
 Let us attempt to determine the
asymptotic behavior of 
 $R^\alpha_\omega \tilde\mu$ $(\lambda\equiv \omega^2)$ 
 from (\ref{(4.3)}) and (\ref{(4.4)}) with 
 $f=\mu$, $k={1\over 2}$, and $\lambda$ and $\omega$ interchanged: 
 \begin{equation}
 R^\alpha_\omega\mu = 2^{-\alpha} R^\alpha_\lambda\mu 
  +\int^\omega_0 J_{{1\over 2},\alpha} (\omega,\tau) 
 R^\alpha_\sigma\mu \,d\tau,
 \label{(4.12)}\end{equation}
 \begin{equation}
J_{{1\over 2},\alpha}(\omega,\tau) 
 = \sum^{\alpha-1}_{j=0} (-1)^{\alpha-j}
\omega^{-j-1} {1\over j!\,(\alpha-1-j)!} {\Gamma\big({j\over 2}
 +{1\over2}\big)\over \Gamma\big({j\over 2}+{1\over 2}-\alpha\big)}
\,\tau^j.
 \label{(4.13)}\end{equation}
Note that all terms in (\ref{(4.13)}) with $j$ odd are equal to 
zero, since ${1\over 2}(j+1)$ is an integer $\le \alpha$. 

Since $\mu$ is of locally bounded variation, it is bounded as
$\lambda\downarrow 0$. 
 Hence $\partial^{-\alpha}_\lambda\mu =O(\lambda^\alpha)$ 
 at small $\lambda$ (see Remark at end of Sec.~\ref{sec5}), so
$R^\alpha_\lambda$ is bounded there. 
 Thus there is no problem with convergence at the lower limit of the 
integral in (\ref{(4.12)}), as a whole. 
 However, some of the individual terms in (\ref{(4.11)}),
  and hence the remainder term by itself, 
 will be singular as $\tau\downarrow 0$ if $\alpha\ge m$.
Therefore, for {\em each\/} integral encountered when 
 (\ref{(4.11)}) is substituted into (\ref{(4.12)}), one must choose 
an appropriate lower limit of integration, $\tau_0\,$. 
 Let $R$ be the part of the total integral thereby omitted;
  thus the integrand of $R$ equals 
 $J_{{1\over2},\alpha} R^\alpha_\sigma\mu(\tau^2)$ 
for small $\tau$, and is redefined as $\tau$ increases --- 
 in such a way that $R$ is convergent and depends on $\omega$ only 
through the factors $\omega^{-j-1}$. 
 Then $R$ will be of the form
\begin{equation}
 R = \sum^{\alpha-1}_{\scriptstyle j=0 \atop \scriptstyle j~{\rm even}}
 Z_j \omega^{-j-1},
 \label{(4.14)}\end{equation}
where the $Z_j$ are constants, which cannot be determined from the 
information in (\ref{(4.11)}) since they are affected by the 
behavior of $R^\alpha_\lambda \mu$ at small $\lambda$. 
 Next consider the part of the integral in (\ref{(4.12)})
associated with the series in (\ref{(4.11)}):
\begin{equation}
 \sum^\alpha_{s=0} a_{\alpha s} \int^\omega_{\tau_0}
\sum^{\alpha-1}_{\scriptstyle j=0 \atop \scriptstyle j~{\rm even}}
(-1)^{\alpha-j} \omega^{-j-1} {1\over j!\,(\alpha-1-j)!}
  {\Gamma\big({j\over2}+{1\over 2}\big)\over 
 \Gamma\big({j\over 2}+ {1\over 2}-\alpha\big)}
\,\tau^{m-s+j}\, d\tau.
 \label{(4.15)}\end{equation}
When each integral in (\ref{(4.15)}) is evaluated,
  the contribution of the lower limit is of the form (\ref{(4.14)}) 
  and may be henceforth counted as a part of $R$. 
 The contribution of the upper limit is proportional to $\omega^{m-s}$,
unless $m-s$ is odd and negative, 
 in which case the term $j=-m+s-1$ yields something proportional to 
$\omega^{m-s}\ln \omega$ and the other values of $j$ yield more 
terms to be absorbed into $R$. 
 Finally, when integrating the
remainder term in (\ref{(4.11)})
  we may assume that $\tau_0$ and $\omega$ are so
large that for some $K$, the integral is less than
\begin{eqnarray}
K \int^\omega_{\tau_0} |J_{{1\over 2},\alpha} (\omega,\tau)|
\tau^{m-\alpha-1} \, d\tau 
&\le &\sum^{\alpha-1} _{\scriptstyle j=0 \atop 
 \scriptstyle j~{\rm even}}
K_j \omega^{-j-1} \int^\omega_{\tau_0} \tau^{m-\alpha+j-1}\, d\tau
 \nonumber\\*
&= &\sum^{\alpha-1}_{\scriptstyle j=0 \atop \scriptstyle j~{\rm even}}
K_j(m-\alpha+j)^{-1} (\omega^{m-\alpha-1} - \omega^{-j-1}
\tau^{m-\alpha+j}_0),
 \label{(4.16)}\end{eqnarray}
except that if $m-\alpha+j=0$, the corresponding term is
\begin{equation}
 K_j \omega^{m-\alpha-1} (\ln \omega - \ln \tau_0).
 \label{(4.17)}\end{equation}
Now $\tau_0$ must be chosen differently for different $j$:
  If $m-\alpha +j<0$, take $\tau_0\to \infty$, leaving in 
 (\ref{(4.16)}) a single contribution to
the error of order $O(\omega^{m-\alpha-1})$. 
 (An integral over a finite interval is thereby included in $R$, 
but (\ref{(4.14)}) is still valid.) 
 If $m-\alpha+j\ge 0$, we have $-j-1 \le m-\alpha-1$, 
 so that {\em both\/} terms in (\ref{(4.16)}) are 
 $O(\omega^{m-\alpha-1})$ for finite $\tau_0\,$, with a
possible extra logarithmic factor in the worst case (\ref{(4.17)}).

Therefore, adding all contributions in (\ref{(4.12)}),
  one has that whenever
$R^\alpha_\lambda\mu$ has the asymptotic behavior (\ref{(4.11)}),
$R^\alpha_\omega\mu$ has the behavior (as $\omega\to\infty$)
\begin{equation}
 R^\alpha_\omega\mu = \sum^\alpha_{s=0} c_{\alpha s}
\omega^{m-s} + \sum^\alpha_{\scriptstyle s=m+1\atop 
 \scriptstyle s-m~{\rm odd}} d_{\alpha s} \omega^{m-s} \ln \omega
+ O(\omega^{m-\alpha-1}\ln \omega),
 \label{(4.18)}\end{equation}
where
\begin{equation}
 c_{\alpha s} \text{ is undetermined if $s>m$ and $s-m$ is
odd;}
 \label{(4.19)}\end{equation}
\begin{eqnarray}
 d_{\alpha s} = (-1)^{\alpha+m-s+1} {1\over (s-m-1)!\,(m-s+\alpha)!}
{\Gamma\big({s\over 2}-{m\over 2}\big)\over \Gamma\big({s\over 2}
 - {m\over2}-\alpha\big)} \,a_{\alpha s}&
\nonumber\\*
\text{(if $s>m$ and $s-m$ is odd);}&
  \label{(4.20)}\end{eqnarray}
\begin{eqnarray}
 c_{\alpha s} = \left[\sum^{\alpha-1}_{\scriptstyle j=0 \atop 
 \scriptstyle j~{\rm even}}(-1)^{\alpha-j}
{(m-s+j+1)^{-1}\over j!\,(\alpha-1-j)!} {\Gamma\big({j\over 2} 
 + {1\over2}\big)\over \Gamma\big({j\over 2} 
 + {1\over 2}-\alpha\big)}\right]
a_{\alpha s} + 2^{-\alpha} a_{\alpha s}&
 \nonumber\\*
\text{if $s\le m$ or $s-m$ is even.}&
 \label{(4.21)}\end{eqnarray}

      \goodbreak
{\em Remark:\/} Writing $\ln \omega$ in (\ref{(4.18)})
  instead of $\ln \kappa\omega$, $\kappa$ some numerical constant,
  is arbitrary.
  In fact, in an
application, $\kappa$ is likely to have physical dimensions 
 (such as length).
Changing $\kappa$ redefines the undetermined coefficient 
 $c_{\alpha s}$ by adding a multiple of $d_{\alpha s}$.

   \paragraph*{\bf 2.}
Now we contrast the foregoing calculation with the 
parallel calculation of $R^\alpha_\lambda\mu$ from 
$R^\alpha_\omega\mu$, 
 assuming
the latter to have the form (\ref{(4.18)}). 
 We use (\ref{(4.3)}) and (\ref{(4.4)}) with $k=2$:
\[
 R^\alpha_{\lambda}\mu = 2^\alpha R^\alpha_\omega\mu +
\int^\lambda_0 J_{2,\alpha}(\lambda,\sigma) R^\alpha_\tau \mu\, 
 d\sigma,
 \]
 \[
J_{2,\alpha}(\lambda,\sigma) 
 = \sum ^{\alpha-1}_{j=0} (-1)^{\alpha-j}
\lambda^{-j-1} {1\over j!\,(\alpha-1-j)!} 
 {\Gamma(2j+2)\over\Gamma(2j+2-\alpha)}\,\sigma^j.
 \]
This time the $j$ term vanishes for all $j\le {1\over 2}\alpha-1$
(equivalently, $j< [\alpha/2]$). 
 Hence the contribution of small $\sigma$ will be of the form 
 (as $\lambda\to\infty$)
 \begin{equation}
 R = \sum^{\alpha-1}_{j=\big[{\alpha\over 2}\big]}
  Z_j \lambda^{-j-1} =
O(\lambda^{-\alpha/2}).
 \label{(4.22)}\end{equation}
The integral of the remainder in (\ref{(4.18)})
 (over large $\sigma$) is less than
\begin{equation}
\sum^{\alpha-1}_{j=\big[{\alpha\over 2}\big]} K_j
\lambda^{-j-1} \int^\lambda_{\sigma_0} \sigma^{{m\over 2} - {\alpha \over
2} + j - {1\over 2}} \ln \sigma \, d\sigma
= \text{ terms of form (\ref{(4.22)}) } + 
 O\big(\lambda^{{m\over 2}-{\alpha \over
2}-{1\over 2}} \ln \lambda\big),
 \label{(4.23)}\end{equation}
except that the second term is 
 $O\big(\lambda^{{m\over 2}-{\alpha\over 2}
 -{1\over 2}}(\ln \lambda)^2\big)$ 
 if the value $j = {1\over 2}(\alpha-m-1)$ occurs. 
Here the relevant integral formulas are
\begin{equation}
 \int\lambda^{p-1} \ln \lambda\, d\lambda =
\lambda^p(p^{-1} \ln \lambda -p^{-2})  + C 
 \quad\text{if $ p\ne 0$}, \qquad
\int \lambda^{-1} \ln \lambda\, d\lambda = {1\over 2} (\ln \lambda)^2 +
C.
 \label{(4.24)}\end{equation}
Since $m\ge 1$, it follows that $R$ and the similar terms in 
(\ref{(4.23)}) are of at least as high order as the last term in 
(\ref{(4.23)}). 
  Hence there are, in effect, {\em no\/} undetermined constants of 
  integration when the calculation 
is done in this direction $(\omega\to\lambda)$. 
 The remaining, explicit integrals are
\begin{eqnarray*}
R_1 &=& \sum^\alpha_{s=0} c_{\alpha s} \int^\lambda_{\sigma_0}
\sum^{\alpha-1}_{j=\big[{\alpha\over 2}\big]} (-1)^{\alpha-j}
\lambda^{-j-1} {1\over j!\,(\alpha-1-j)!} {\Gamma(2j+2)\over
\Gamma(2j+2-\alpha)} \,\sigma^{{m\over 2}-{s\over 2}+j}\, d\sigma\\*
&&{} + \sum^\alpha_{\scriptstyle s=m+1\atop \scriptstyle s-m~{\rm odd}}
{1\over 2}d_{\alpha s} \int^\lambda_{\sigma_0}
\sum^{\alpha-1}_{j=\big[{\alpha\over 2}\big]} (-1)^{\alpha-j}
\lambda^{-j-1} {1\over j!\,(\alpha-j)!} {\Gamma(2j+2)\over
\Gamma(2j+2-\alpha)} \,\sigma^{{m\over 2}-{s\over 2}+j} \ln \sigma\,
d\sigma.
 \end{eqnarray*}
Using (\ref{(4.24)}),
  one sees that the values of $s$ for which $d_{\alpha s} = 0$
yield terms proportional to $c_{\alpha s} \lambda^{(m-s)/2}$,
  but that when
$d_{\alpha s}\ne 0$ there are, {\em a priori}, two types of term,
proportional to
\[
 c_{\alpha s} \lambda^{{m\over 2}-{s\over 2}}
  + {1\over 2} d_{\alpha s}\lambda^{{m\over 2}-{s\over 2}} \ln \lambda 
 \quad \text{and}\quad 
 d_{\alpha s} \lambda^{{m\over 2}-{s\over 2}},
\]
respectively. (Since
\[ {m\over 2}-{s\over 2} + j > -{\alpha\over 2}	+ \left({\alpha\over
2}-1\right) = -1,
\]
no logarithms are ``created'' by the integration as in (\ref{(4.15)}).)
  Since it must be possible to recover the expansion (\ref{(4.11)}) 
  in this way, there must be a numerical coincidence which causes 
  all terms proportional to $d_{\alpha s} \ln \lambda$ to cancel. 
 By the observation
just made, this same numerical identity will cause all terms involving
$c_{\alpha s}$ to cancel, if $s-m$ is odd and positive. 
 Thus these numbers
$c_{\alpha s}$ do not affect at all the asymptotic behavior of
$R^\alpha_\lambda\mu$ --- 
 as was to be expected from the fact that the latter
does not determine them (\ref{(4.19)}).
  Obviously, similar cancellations of
logarithmic and $c_{\alpha s}$ terms must occur 
 when $R^\alpha_\lambda \mu$
is calculated from $R^\alpha_\omega\mu,\ldots, R^{2\alpha}_\omega\mu$ by
H\"or\-man\-der's formula (\ref{(4.10)}).

The conclusion therefore is that if $R^\alpha_\omega\mu$ has the 
asymptotic behavior (\ref{(4.18)}), then $R^\alpha_\lambda\mu$ has the 
behavior (\ref{(4.11)}) with error term 
$ O\bigl(\lambda^{{m\over 2}-{\alpha\over 2}-{1\over 2}} (\ln
\lambda)^2\bigr)$
in the worst case. The formulas for the coefficients are
\begin{eqnarray}
 a_{\alpha s} = \left[-{1\over 2} \sum^{\alpha-1}_{j=\big[{\alpha\over
2}\big]} (-1)^{\alpha-j} {\big({m\over 2}-{s\over 2} + j+1)^{-2} \over
j!\,(\alpha-1-j)!} {\Gamma(2j+2)\over \Gamma(2j+2-\alpha)}\right]\,
  d_{\alpha s}&
\nonumber\\*
\text{if $s>m$ and $s-m$ is odd;}&
  \label{(4.25)}\end{eqnarray}
\begin{eqnarray}
 a_{\alpha s} = \left[\sum^{\alpha-1}_{j=\big[{\alpha\over 2}\big]}
(-1)^{\alpha-j} {\big({m\over 2}-{s\over 2} +j+1\big)^{-1} \over
j!\,(\alpha-1-j)!} {\Gamma(2j+2)\over \Gamma(2j+2-\alpha)}\right]\, 
 c_{\alpha s} + 2^\alpha c_{\alpha s}&
 \nonumber\\*
 \text{if $s\le m$ or $s-m$ is even.}&
  \label{(4.26)}\end{eqnarray}
\goodbreak

The conditions of consistency between the two calculations are 
 (for $0 \le s \le \alpha$, $m\ge 1$, $\alpha\ge 1$)
\begin{eqnarray}
\left[-{1\over 2}\sum^{\alpha-1}_{j=\big[{\alpha\over
2}\big]} (-1)^{\alpha-j} 
 {\big({m\over 2} -{s\over 2} + j+1\big)^{-2} \over
j!\,(\alpha-1-j)!} {\Gamma(2j+2)\over
\Gamma(2j+2-\alpha)}\right]\qquad&
 \nonumber\\*
\quad \times \left[(-1)^{\alpha+m-s+1} {1\over (s-m-1)!\,(m-s+\alpha)!}
{\Gamma\big({s\over 2}-{m\over 2}\big)\over \Gamma\big({s\over 2} 
 - {m\over2}-\alpha\big)}\right] = 1&
 \nonumber\\*
\hfil \text{if $s>m$ and $s-m$ is odd}&
  \label{(4.27)}\end{eqnarray}
(from (\ref{(4.20)}) and (\ref{(4.25)}));
\begin{eqnarray}
\left[\sum^{\alpha-1}_{j=\big[{\alpha\over 2}\big]}
(-1)^{\alpha-j} {\big({m\over 2}-{s\over 2} +j+1\big)^{-1}
  \over j!\,(\alpha-1-j)!} {\Gamma(2j+2)\over \Gamma(2j+2-\alpha)} 
 + 2^\alpha\right]\qquad&
\nonumber\\*
\quad \times \left[\sum^{\alpha-1}_{\scriptstyle j=0\atop
  \scriptstyle j\text{ even}}(-1)^{\alpha-j} {(m-s+j+1)^{-1}
\over j!\,(\alpha-1-j)!} {\Gamma\big({j\over 2} 
 + {1\over 2}\big) \over\Gamma \big({j\over 2}
 + {1\over 2}-\alpha\big)} + 2^{-\alpha}\right] =1&
\nonumber\\*
\hfil \text{if $s\le m$ or $s-m$ is even}&
\label{(4.28)}\end{eqnarray}
(from (\ref{(4.21)}) and (\ref{(4.26)}));
\begin{eqnarray}
 \sum^{\alpha-1}_{j=\big[{\alpha\over 2}\big]} (-1)^{\alpha-j}
{\big({m\over 2} - {s\over 2}+j+1\big)^{-1}\over j!\,(\alpha-1-j)!}
{\Gamma(2j+2)\over \Gamma(2j+2-\alpha)} + 2^\alpha = 0&
 \nonumber\\*
 \text{if $s>m$ and $s-m$ is odd}&
  \label{(4.29)} \end{eqnarray}
(the ``numerical coincidence''). These can easily be verified for small
values of $\alpha$.

In the Appendix  it is shown that (\ref{(4.28)}) is valid for 
all values of $s-m$ as a complex variable, except those where one 
of the factors is undefined. 
 Then (\ref{(4.29)}) is obtained (for any $\alpha$) in the limit as 
 $s-m$ approaches a pole of the second factor, 
 and (\ref{(4.27)}) is obtained similarly from the
derivative of (\ref{(4.28)}).

\section{The means of a Stieltjes integral in terms of the
means of the measure}\label{sec5}

The calculation in this section is the central lemma relating the 
asymptotics of Green functions at small $t$ to the asymptotics of 
spectral measures of various kinds at large $\lambda$. 

Let $\mu$ and $f$ be functions of locally bounded variation, 
vanishing at 0, related as in (\ref{(3.2)}):
  \[ f(\lambda) = \int^\lambda_0 
a(\sigma)\, d\mu(\sigma). \] 
 Assume that $a(\lambda)$ is a $C^\infty$  
function, and that it is well-behaved at the origin as described in 
the remark at the end of this section. 
 Integrating by parts, one 
has \begin{equation} 
 f(\lambda) = a(\lambda) \mu(\lambda) - \int^\lambda_0 a'(\sigma)
\mu(\sigma) \,d\sigma.
 \label{(5.1)}\end{equation}

Continuing similarly, one can express 
 $\partial^{-\alpha}_\lambda f$ in terms of integrals whose 
integrands involve only $\partial^{-\alpha}_\lambda\mu$ and 
derivatives of $a$: 
 \begin{eqnarray}    
\partial^{-\alpha}_\lambda f(\lambda) &=& a(\lambda)
\partial^{-\alpha}_\lambda\mu(\lambda)
 + \sum^{\alpha+1}_{j=1} (-1)^j {\alpha+1\choose j} {1\over (j-1)!}
\int^\lambda_0 (\lambda-\sigma)^{j-1} \partial^j_\sigma a(\sigma)
\partial^{-\alpha}_\sigma \mu(\sigma)\,d\sigma
 \nonumber\\*
&=& \sum^{\alpha+1}_{j=0} \int^\lambda_0 d\sigma_1\cdots
\int^{\sigma_{j-1}}_0 d\sigma\, 
(-1)^j {\alpha+1 \choose j} \partial^j_\sigma
a(\sigma) \partial^{-\alpha}_\sigma\mu (\sigma).
 \label{(5.2)}\end{eqnarray}
(The integral in the last version is $j$-fold.) This relation can be
written
\begin{equation}
R^\alpha_\lambda f(\lambda) = a(\lambda) R^\alpha_\lambda
\mu(\lambda)
 + \lambda^{-\alpha} \sum^{\alpha+1}_{j=1} (-1)^j {\alpha+1\choose j}
{1\over (j-1)!} \int^\lambda_0 (\lambda-\sigma)^{j-1} \sigma^\alpha
\partial^j_\sigma a(\sigma) R^\alpha_\sigma\mu(\sigma)\,d\sigma.
 \label{(5.3)}\end{equation}

{\em Proof:\/}
  For $\alpha=0$, (\ref{(5.2)}) is (\ref{(5.1)}).
  Assume (\ref{(5.2)}) for $\alpha$ and prove it for $\alpha+1$:  
  First, 
 \begin{eqnarray*}
 \partial^{-\alpha-1}_\lambda f(\lambda) &=& \int^\lambda_0
\partial^{-\alpha}_\sigma f(\sigma)d\sigma
 \\* &=& \int^\lambda_0 a(\sigma)
\partial^{-\alpha}_\sigma \mu(\sigma)d\sigma
 +\sum^{\alpha+1}_{j=1} (-1)^j {\alpha+1\choose j} \int^\lambda_0
d\sigma \int^\sigma_0 d\sigma_1\cdots \int^{\sigma_{j-1}}_0 d\tau\,
\partial^j_\tau a(\tau)\partial^{-\alpha}_\tau \mu(\tau).
\end{eqnarray*} 
The last term can be rewritten by (\ref{(3.6)}) as
\[ \sum^{\alpha+1}_{j=1} (-1)^j {\alpha+1\choose j} {1\over j!}
\int^\lambda_0 (\lambda-\sigma)^j \partial^j_\sigma a(\sigma)
\partial^{-\alpha}_\sigma\mu(\sigma)\,d\sigma.
\]
Now integrate by parts in both terms:
\begin{eqnarray*}
\partial^{-\alpha-1}_\lambda f(\lambda) &=& a(\lambda)
\partial^{-\alpha-1}_\lambda \mu(\lambda) 
 - \int^\lambda_0 \partial_\sigma
a(\sigma) \partial^{-\alpha-1}_\sigma \mu(\sigma)\,d\sigma
 \\*&&{}
  + \sum^{\alpha+1}_{j=1} (-1)^{j+1} {\alpha+1\choose j} 
 {1\over  j!}\int^\lambda_0 
 \partial_\sigma[(\lambda-\sigma)^j \partial^j_\sigma a(\sigma)]
  \partial^{-\alpha-1}_\sigma \mu(\sigma)\,d\sigma.
 \end{eqnarray*}
The last term equals
\begin{eqnarray*}
&&\sum^{\alpha+1}_{j=1} (-1)^j {\alpha+1\choose j} {1\over
(j-1)!} \int^\lambda_0 (\lambda-\sigma)^{j-1} 
 \partial ^j_\sigma a(\sigma)
\partial ^{-\alpha-1}_\sigma \mu(\sigma)\,d\sigma\\*
&+&\sum^{\alpha+2}_{j=2} (-1)^j {\alpha+1\choose j-1} {1\over (j-1)!}
\int^\lambda_0 (\lambda-\sigma)^{j-1} \partial^j_\sigma a(\sigma)
\partial^{-\alpha-1}_\sigma\mu(\sigma)\,d\sigma.
 \end{eqnarray*}
Using Pascal's triangle relation
\begin{equation}
 {\alpha+1\choose j} + {\alpha+1\choose j-1} = {\alpha+2\choose j}, 
 \label{(5.4)}\end{equation}
one simplifies the expression to
\[
\partial^{-\alpha-1} _\lambda f(\lambda) 
 = a(\lambda)\partial^{-\alpha-1}_\lambda \mu(\lambda)
+ \sum^{\alpha+2}_{j=1} (-1)^j {\alpha+2\choose j} {1\over (j-1)!}
\int^\lambda_0 (\lambda-\sigma)^{j-1} \partial^j_\sigma a(\sigma)
\partial^{-\alpha-1}_\sigma\mu(\sigma)\,d\sigma,
 \]
as was to be proved.


{\em Remark:\/} In the foregoing it was tacitly assumed that 
$\partial^j_\sigma a(\sigma)$ was bounded as $\sigma\downarrow 0$, 
so that there were no lower-endpoint contributions in the 
integrations by parts. 
 Actually, since 
 \begin{equation}
 \partial^{-\alpha}_\sigma\mu(\sigma) = O(\sigma^\alpha) \quad\text{as 
$\sigma\downarrow 0$}
 \label{(5.5)}\end{equation}
(as follows from (\ref{(3.6)}) and the boundedness of $\mu$), it 
suffices to assume that 
 \begin{equation}
 \partial^\alpha_\sigma a(\sigma) = o(\sigma^{-\alpha}) \quad\text{as 
$\sigma\downarrow 0$}.
 \label{(5.6)}\end{equation}

\section{Heat kernels (Laplace transforms) and Riesz means}
 \label{sec6}

 The standard heat kernel of a second-order operator has a natural 
association with the variable we have called~$\lambda$,
 while the cylinder kernel (introduced in Sec.~\ref{sec2}) is 
associated in the same way with the variable~$\omega$.
 The terms in the asymptotic expansions of these Green functions 
are in direct correspondence with those in the asymptotic 
expansions of the associated Riesz means.
 However, it is instructive to try to calculate each Green 
function in terms of the ``wrong'' variable,
 to observe how information gets lost, or needs to be resupplied,
in passing from one quantity to another.
 Therefore, this section divides naturally into four parts.

 \paragraph*{\bf 1.}
Given $\mu(\lambda)$, a function of locally bounded variation on
$[0,\infty)$, we consider
\begin{equation}
 K(t) \equiv \int^\infty_0 e^{-\lambda t} d\mu(\lambda)\qquad (t>0).
 \label{(6.1)} \end{equation} 
 As $t\downarrow 0$, we anticipate an expansion of the form
\begin{equation}
 K(t) \sim \sum^\infty_{s=0} b_s t^{-{m\over 2}+{s\over 2}};
 \label{(6.2)} \end{equation}
we shall demonstrate detailed equivalence of (\ref{(6.2)}) with 
 (\ref{(4.11)}).

In spectral theory, (\ref{(6.1)}) has a number of possible 
 interpretations.
  Let $H$
be a positive, elliptic, second-order differential operator on an
$m$-dimensional manifold. 
 Then: (1) If the manifold is compact, $K$ may be the integrated 
``trace'' of the heat kernel of $H$, $\mu(\lambda)$ being the 
number of eigenvalues less than or equal to $\lambda$. 
 (2) $K$ may be the diagonal value of the heat kernel at a point 
$x$, $\mu(\lambda) \equiv E_\lambda(x,x)$ being the diagonal value 
of the spectral function (integral kernel of the spectral 
projection). 
 In this case one knows that $b_s=0$ if $s$ is odd. 
 (3) $K$ may be the diagonal value of some spatial derivative of the 
heat kernel, $\mu$ being the corresponding derivative of the 
spectral function. 
 (In this case, $m$ in (\ref{(6.2)}) depends on the order of the 
derivative as well as on the dimension.) 
 If $m=1$, $\mu$ is a Titchmarsh-Kodaira spectral measure. 
 (4) $K = K(t,x,y)$ may be the full heat kernel (off-diagonal), 
$\mu$ being the full spectral function. 

Let us assume (\ref{(4.11)}),
\[ R^\alpha_\lambda\mu(\lambda) 
 = \sum^\alpha_{s=0} a_{\alpha s}
\lambda^{{m\over 2}-{s\over 2}} + O(\lambda^{{m\over 2}-{\alpha\over 2}-
{1\over 2}}), \]
and calculate $K(t)$ from (\ref{(5.3)}),
  with $a(\lambda) = e^{-\lambda t}$,
\[ \partial^j_\lambda a(\lambda)  = (-t)^j e^{-\lambda t}, \]
and $\lambda\to \infty$ in (\ref{(5.3)}).
  (See the first theorem of consistency, (\ref{(3.13)}).)
  In the present case all integrals in (\ref{(5.3)}) converge 
 as $\lambda\to\infty$, 
 and hence the only term that survives in the limit is the one
with $j=\alpha+1$, $(\lambda-\sigma)^{j-1} = \lambda^\alpha
+o(\lambda^\alpha)$. 
 So
 \begin{eqnarray*} 
 K(t) &=& (-1)^{\alpha+1} {1\over \alpha!} \int^\infty_0
\sigma^\alpha(-t)^{\alpha+1} e^{-\sigma t}
R^\alpha_\sigma\mu(\sigma)\,d\sigma\\*
&=& t^{\alpha+1} {1\over \alpha!} \int^\infty_0 
 \left[\sum^\alpha_{s=0}
a_{\alpha s} \sigma^{{m\over 2}-{s\over 2}+\alpha} 
 + O (\sigma^{{m\over 2}
+ {\alpha\over 2}-{1\over 2}})\right] e^{-\sigma t} d\sigma. 
 \end{eqnarray*}
Using
 \begin{equation}
 \int^\infty_0 \lambda^{p-1} e^{-\lambda t} d\lambda 
 = \Gamma(p) t^{-p}\qquad (p>0),
 \label{(6.3)} \end{equation}
we have
\begin{equation}
 K(t) = \sum^\alpha_{s=0} 
 {\Gamma\big({m\over 2}-{2\over 2}+\alpha+1\big)\over
  \Gamma(\alpha+1)} a_{\alpha s} t^{-{m\over 2}+{s\over 2}} +
O(t^{-{m\over 2}+{\alpha\over 2}+{1\over 2}}).
 \label{(6.4)} \end{equation}
(The undeterminable contribution from small $\sigma$ is 
 $O(t^{\alpha+1})$,
which is of higher order than the remainder.) 
 Thus
\begin{equation}
 b_s = {\Gamma\big({m\over 2}-{s\over 2} + \alpha+1\big) \over
\Gamma(\alpha+1)}\, a_{\alpha s}
 \label{(6.5)} \end{equation}
holds for $\alpha \ge s$, and formally for smaller $\alpha$. 
 (In the latter context a pole of 
 $\Gamma\big({1\over 2}(m-s)+\alpha+1\big)$ may be
encountered. 
 The significance of (\ref{(6.5)}) then is that $a_{\alpha s}=0$. 
 Thus, when $m-s$ is even, $\alpha$ must be sufficiently large 
before $a_{\alpha s}$ will uniquely determine $b_s\,$, even 
formally.) 
 For $a<s$, (\ref{(6.5)}) may be
taken as a definition of $a_{\alpha s}\,$.

To check that the right-hand side of (\ref{(6.5)})
  is independent of $\alpha$,
write (\ref{(4.11)}) as
\begin{equation}
 \partial^{-\alpha} _\lambda\mu(\lambda) \sim {1\over \alpha!}
\sum^\alpha_{s=0} a_{\alpha s} \lambda^{{m\over 2}-{s\over 2}+\alpha}
 \label{(6.6)} \end{equation}
and differentiate formally:
\[
 \partial^{-\alpha+1}_\lambda\mu(\lambda) \sim {1\over \alpha!} 
 \sum_s
\left({m\over 2}-{s\over 2}+\alpha\right) a_{\alpha s} 
 \lambda^{{m\over2}-{s\over 2} +\alpha-1}. \]
By comparison with (\ref{(6.6)}) for $\alpha-1$, one has
\[
 {1\over\alpha!} \left({m\over 2}-{s\over 2}+\alpha\right) 
 a_{\alpha s} =
{1\over (\alpha-1)!} a_{\alpha-1,s}\,, \]
which may be variously rewritten as
\begin{equation}
 {\Gamma\big({m\over 2}-{s\over 2}+\alpha+1\big)\over 
 \Gamma(\alpha+1)}
a_{\alpha s} = {\Gamma\big({m\over 2}-{s\over 2}+\alpha\big)\over
\Gamma(\alpha)} a_{\alpha-1,s}
 \label{(6.7)} \end{equation}
(showing consistency of (\ref{(6.5)})), or as
\begin{equation}
 a_{\alpha-1,s} = {1\over \alpha}\left({m\over 2} 
 - {s\over 2}+\alpha\right)a_{\alpha s}\qquad (\alpha\ge 1).
 \label{(6.8)} \end{equation}
In particular, we can stem the profusion of constants by choosing 
 $a_{ss}$
as the fiducial member of the family $\{a_{\alpha s}\}$.
  We have
\begin{equation}
 b_s = {\Gamma\big({m\over 2} + {s\over 2}+1\big)\over \Gamma(s+1)}
\,a_{ss}\,, 
 \label{(6.9)} \end{equation}
and
\begin{equation}
 a_{\alpha s} = {\Gamma(\alpha+1) \Gamma\big({m\over 2}
 +{s\over 2} +1\big)
\over \Gamma\big({m\over 2}  - {s\over 2} +\alpha+1\big) \Gamma(s+1)}
\,a_{ss} 
 \label{(6.10)} \end{equation}
(meaning 0, of course, when the first factor in the denominator has 
a pole).

{\em Remark:\/}
  When $m-s$ is even, $a_{\alpha s}$ is zero for small 
$\alpha$, and the nonzero value for large $\alpha$ 
 (hence the nonzero value of $b_s$) 
 may be regarded as a constant of integration encountered in the 
passage from a low-order Riesz mean to a higher-order one. 
 A case of particular interest is the local diagonal value of the 
heat kernel, for which $b_s=0$ when $s$ is odd but generally 
$b_s\ne 0$ for $s$ even. 
 Here there is an essential difference between $m$ even and $m$ 
odd. 
 If $m$ is odd, then $a_{0s}$ is 0 for odd $s$ and nonzero for even 
$s$, and the constants of integration are also~$0$. 
 If $m$ is even, then $a_{0s} = 0$ for {\em all\/} $s>m$, and the 
nonzero values of $b_s$ for $s>m$, $s$ even, come entirely from 
constants of integration. 
 This dimensional property of $a_{0s}$ is reflected in the poles of 
the zeta function.\cite{MP,DG}

 \paragraph*{\bf 2.}
It is of interest to consider how (\ref{(6.2)}) could be calculated 
from the $\omega$-means, (\ref{(4.18)}). 
  We have
\begin{eqnarray}
 \partial_\omega(e^{-\omega^2t}) &=& e^{-\omega^2t}
(-2\omega t),\nonumber\\*
\partial^2_\omega(e^{-\omega^2t}) &=& e^{-\omega^2t} (-2t +
4\omega^2t^2),\nonumber\\*
\partial ^3_\omega(e^{-\omega^2t}) &=& e^{-\omega^2t} (12 \omega t^2 -
8\omega^3t^3),\nonumber\\*
\partial^4_\omega(e^{-\omega^2t}) &=& e^{-\omega^2t} (12 t^2 - 48
\omega^2t^3 + 16 \omega^4 t^4),
 \label{(6.11)} \end{eqnarray}
and in general the form
\begin{equation}
 \partial^j_\omega(e^{-\omega^2t}) 
 = e^{-\omega^2t} \sum^{\big[{j\over
2}\big]}_{i=0} z_i \omega^{j-2i} t^{j-i}.
 \label{(6.12)} \end{equation}
As before, when $\omega\to \infty$ the formula (\ref{(5.3)})
  reduces to one term,
\[
 K(t) = \lim_{\omega\to \infty} R^\alpha_\omega\left[
\int^\omega_0 e^{-\tau^2t} d\mu(\tau)\right]
= (-1)^{\alpha+1} {1\over \alpha!} \int^\infty_0 \omega^\alpha
\partial^{\alpha+1}_\omega (e^{-\omega^2t}) R^\alpha_\omega\mu\, d\omega
 ,\]
where (\ref{(4.18)})
\[
 R^\alpha_\omega\mu = \sum^\alpha_{s=0} c_{\alpha s} \omega^{m-s} +
\sum^\alpha_{\scriptstyle s=m+1\atop \scriptstyle s-m\text{ odd}} 
 d_{\alpha s} \omega^{m-s} \ln \omega 
 + O(\omega^{m-\alpha-1} \ln \omega).
  \]
The relevant integrals this time are $(p>0)$
\begin{equation}
 \int^\infty_0 e^{-\omega^2t} \omega^{2p-1} d\omega = {1\over
2} \Gamma(p)t^{-p},
 \label{(6.13)}\end{equation}
 \nobreak
\begin{equation}
 \int^\infty_0 e^{-\omega^2t} \omega^{2p-1} \ln \omega \, d\omega 
 = {1\over4} \Gamma(p) [\psi(p) - \ln t]t^{-p},
\label{(6.14)} \end{equation}
where
\begin{equation}
 \psi(p) \equiv \partial_p \ln \Gamma(p)
 \label{(6.15)} \end{equation}
satisfies
\begin{equation}
 \psi(p+1) = \psi(p) + {1\over p}\,,
 \label{(6.16)}\end{equation}
 \begin{equation}
\psi(n) = -\gamma + \sum^{n-1}_{k=1} {1\over k}\qquad
(n=1,2,\ldots),\label{(6.17)}\end{equation}
 \begin{equation}
\psi\left({1\over 2}+n\right) 
 = -\gamma -2\ln 2 + 2 \sum^n_{k=1} {1\over 2k-1} \qquad 
 (n=0,1,\ldots), 
 \label{(6.18)}\end{equation}
 \begin{equation}
\gamma \equiv +0.57721566490 \qquad \text{(Euler's constant)}.
 \label{(6.19)} \end{equation}
 From (\ref{(6.12)}) we see that the contribution of small $\omega$ 
is 
 $
 O\bigl(t^{[(\alpha+2)/2]}\bigr) 
$,
and the integral of the error term out to $\omega=\infty$ is
$
 O\bigl(t^{-{m\over 2}+{\alpha\over 2}+{1\over 2}} \ln t\bigr)
 $;
the latter is dominant, as in (\ref{(6.4)}).
  Therefore, the remaining, explicit
terms should reproduce exactly the summation in (\ref{(6.4)}).
  This requires cancellation of all $\ln t$ terms; 
 comparison of (\ref{(6.13)}) with (\ref{(6.14)}) shows that this 
will entail cancellation of all $c_{\alpha s}$ terms with $s-m$ odd 
and positive, 
 just as in the calculation at the end of Sec.~\ref{sec4}
   (and also, coincidentally, of all terms containing $\gamma+2 \ln 2$). 
  We illustrate by working out the case $m=1$, $\alpha=3$, using 
  the last of Eqs.~(\ref{(6.11)}):
 \begin{eqnarray*}
K(t) &\sim& {1\over 6} \int^\infty_0 e^{-\omega^2t} (12 t^2-48
\omega^2 t^3 + 16 \omega^4t^4)\\*
 &&{} \times [c_{30} \omega^4 + c_{31} \omega^3 + c_{32}\omega^2 + 
c_{33} \omega + d_{32} \omega^2 \ln \omega]d\omega\\ 
 &=& c_{30} t^{-1/2} \textstyle\left[\Gamma\left({5\over 2}\right)
-4\Gamma\left({7\over 2}\right) + {4\over 3}\Gamma\left({9\over
2}\right)\right] + c_{31}\left[\Gamma(2) - 4\Gamma(3) + {4\over 3}
\Gamma(4)\right]\\*
&&{} + c_{32} t^{1/2} \textstyle\left[\Gamma\left({3\over 2}\right) 
-4\Gamma\left({5\over 2}\right) + {4\over 3}\Gamma\left({7\over 
2}\right) \right] + c_{33} t\left[\Gamma(1) - 4\Gamma(2) + {4\over 
3} \Gamma(3)\right]\\* 
 &&{} -{1\over 2} d_{32}(\gamma + 2\ln 2 + \ln t)t^{1/2} 
\textstyle\left[\Gamma
\left({3\over 2}\right) - 4\Gamma\left({5\over 2}\right) +
{4\over3}\Gamma\left({7\over 2}\right)\right]\\*
&&{}+ d_{32} t^{1/2} \textstyle\left[\Gamma\left({3\over 2}\right) 
- {16\over 3} \Gamma\left({5\over 2}\right) + {92 \over 45} 
\Gamma\left({7\over 2}\right) \right]\\*
 &=& \textstyle 
2\Gamma\left({1\over 2}\right) c_{30}t^{-1/2} + c_{31} + {1\over 3} 
\Gamma\left({1\over 2}\right) d_{32} t^{1/2} - {1\over 3} 
c_{33}t.
 \end{eqnarray*}
  Moreover, the numerical coefficients 
agree with those computed from (\ref{(6.4)})--(\ref{(6.5)})
  and (\ref{(4.25)})--(\ref{(4.26)}). 

 \paragraph*{\bf 3.}
 Next consider the quantity
\begin{equation}
 T(t) \equiv \int^\infty_0 e^{-\omega t} d\mu\qquad (t>0),
 \label{(6.20)} \end{equation}
where $\mu = \mu(\lambda) = \mu(\omega^2) = \tilde\mu(\omega)$. 
 $K(t)$ bears the same relation to the heat kernel of the operator 
$H$ which $T(t)$ bears to the kernel of the operator 
 $\exp(-\sqrt H\, t)$, the cylinder kernel.
 That operator solves the elliptic partial differential equation 
 \begin{equation}
 \partial^2_t\psi(t,x) - H\psi(t,x) = 0\qquad (t>0)
 \label{(6.21)} \end{equation}
in a cylindrical manifold of dimension $m+1$, with inhomogeneous 
Dirichlet data on the $m$-dimensional boundary surface $t=0$ and a 
decay condition as $t\to\infty$. 

The calculation of the small-$t$ expansion of $T(t)$ from the 
$\omega$-mean expansion (\ref{(4.18)}) starts off in precise 
analogy to the previous calculation of the expansion of $K(t)$ from 
the $\lambda$-mean expansion (\ref{(4.11)}). 
 One obtains from (\ref{(5.3)}), with $\omega$ in the role of 
$\lambda$, 
 \begin{eqnarray*}
 T(t) &=& (-1)^{\alpha+1} {1\over 
\alpha!} \int^\infty_0 \tau^\alpha (-t)^{\alpha+1} e^{-\tau t} 
R^\alpha_\tau \tilde \mu\, d\tau\\*
  &=& t^{\alpha+1} {1\over \alpha!} \int^\infty_0 
 \Bigg[\sum^\alpha_{s=0} c_{\alpha s} 
\omega^{m-s+\alpha} + \sum^\alpha_{\scriptstyle s=m+1 \atop 
\scriptstyle s-m\text{ odd}} d_{\alpha s} \omega^{m-s+\alpha} \ln 
\omega  +O(\omega^{m-1}\ln \omega)\Bigg]  e^{-\omega t} d\omega.
  \end{eqnarray*}
In addition to (\ref{(6.3)}) in the form
\[
 \int^\infty_0 \omega^{p-1} e^{-\omega t} d\omega 
 = \Gamma(p)t^{-p}
 \]
we need a transformation of (\ref{(6.14)}),
\begin{equation}
 \int^\infty_0 e^{-\omega t} \omega^{p-1} \ln \omega \, d\omega = 
\Gamma(p) [\psi(p) - \ln t]t^{-p}\qquad (p>0). 
 \label{(6.22)} \end{equation}
The integral of the error term is of order $O(t^{-m+\alpha+1}\ln 
t)$, and the contribution from small $\omega$ is $O(t^{\alpha+1})$, 
hence of higher order than the error term. 
 Evaluating the integrals of the summations, we get 
 \begin{eqnarray}
T(t) &=& \sum^\alpha_{s=0} {\Gamma(m-s+\alpha+1)\over
\Gamma(\alpha+1)} \,c_{\alpha s}t^{-m+s}
 \nonumber\\*
&&{} + \sum^\alpha_{\scriptstyle s=m+1\atop \scriptstyle 
 s-m\text{ odd}} {\Gamma(m-s+\alpha+1)\over \Gamma(\alpha+1)} 
[\psi(m-s+\alpha+1) - \ln t] \,d_{\alpha s} t^{-m+s} 
\nonumber\\*
 &&{} + O(t^{-m+\alpha+1} \ln t).
 \label{(6.23)}\end{eqnarray}
Thus, if we define notation by
\begin{equation}
 T(t) \sim \sum^\infty_{s=0} e_s t^{-m+s} + \sum^\infty_{\scriptstyle 
s=m+1\atop \scriptstyle s-m\text{ odd}} f_s t^{-m+s} \ln t, 
 \label{(6.24)} \end{equation}
we have
\begin{equation}e_s = {\Gamma(m-s+\alpha+1)\over \Gamma(\alpha+1)} 
\,c_{\alpha s}
 \label{(6.25)} \end{equation} 
 if $s-m$ is even or negative, and
\begin{equation}
 e_s = {\Gamma(m-s+\alpha+1)\over \Gamma(\alpha+1)} [c_{\alpha
s} + \psi(m-s+\alpha +1)d_{\alpha s}],
 \label{(6.26)}\end{equation}
 \nobreak
\begin{equation}
 f_s = -\,{\Gamma(m-s+\alpha+1)\over \Gamma(\alpha+1)}\,d_{\alpha s}
 \label{(6.27)} \end{equation}
for $s-m$ odd and positive. 
 These equations are rigorously valid for $\alpha\ge s$, and hold 
formally for smaller $\alpha$ --- 
 that is, they can be used to {\em define\/} $c_{\alpha s}$ and 
$d_{\alpha s}$ for $\alpha<s$, with one exception: 
  When
\begin{equation}
 m-s+\alpha < 0,
 \label{(6.28)} \end{equation}
the $\Gamma$ function in the numerator has a pole. 
 In the context of (\ref{(6.25)}) or (\ref{(6.27)}), 
  this is understood to force $c_{\alpha s}=0$ or $d_{\alpha s}=0$.
 However, one may not conclude that the $c_{\alpha s}$ in 
(\ref{(6.26)}) is zero in this situation, since $\psi$ also has a 
pole; 
 indeed, we know that
$c_{\alpha s}$ is generally nonzero in the case $\alpha=0$, $m$ 
odd, where the relation of $R^\alpha_\omega\mu$ to 
$R^\alpha_\lambda\mu$ and the heat kernel $K(t)$ is trivial. 

The right-hand sides of (\ref{(6.25)})--(\ref{(6.27)}) 
 must be independent of $\alpha$. 
 To verify this, rewrite (\ref{(4.18)}) as
\begin{equation}
 \partial^{-\alpha}_\omega\mu \sim {1\over \alpha!} 
\sum^\alpha_{s=0} c_{\alpha s} \omega^{m-s+\alpha} 
 + {1\over \alpha!} \sum^\alpha_{\scriptstyle s=m+1
\atop \scriptstyle s-m\text{ odd}} d_{\alpha s} 
 \omega^{m-s+\alpha} \ln \omega 
 \label{(6.29)} \end{equation}
and differentiate:
\begin{eqnarray*}
\partial^{-\alpha+1}_\omega\mu &\sim& {1\over \alpha!} \sum_s
(m-s+\alpha) c_{\alpha s} \omega^{m-s+\alpha-1}\\*
&&{} + {1\over \alpha!} \sum_{\scriptstyle s\ge m+1\atop 
 \scriptstyle s-m\text{ odd}}d_{\alpha s} \omega^{m-s+\alpha-1} 
 + {1\over \alpha!} \sum_{\scriptstyle s\ge m+1
\atop \scriptstyle s-m\text{ odd}}  (m-s+\alpha) d_{\alpha s}
\omega^{m-s+\alpha-1} \ln \omega.
 \end{eqnarray*}
This yields the recursion relations
\begin{equation}
 {1\over \alpha!} (m-s+\alpha) c_{\alpha s} = {1\over
(\alpha-1)!} \,c_{\alpha-1,s} 
\label{(6.30)} \end{equation}
 for  $s-m$  even or negative;
\begin{equation}
 {1\over \alpha !} (m-s+\alpha) c_{\alpha s} + {1\over \alpha!}\, 
d_{\alpha s} = {1\over (\alpha-1)!} \,c_{\alpha-1,s}\,,
 \label{(6.31)}\end{equation}
 \nobreak
\begin{equation}
 {1\over \alpha!} (m-s+\alpha)d_{\alpha s} = {1\over (\alpha-1)!}
\,d_{\alpha-1,s} 
 \label{(6.32)} \end{equation}
for  $s-m$  odd and positive.
Multiplying by $\Gamma(m-s+\alpha)$, we immediately notice 
consistency of (\ref{(6.30)}) and (\ref{(6.32)}) with 
(\ref{(6.25)}) and (\ref{(6.27)}) [cf.~(\ref{(6.7)})], while 
(\ref{(6.31)}) becomes 
 \[
 {\Gamma(m-s+\alpha+1)\over \Gamma(\alpha+1)} \,c_{\alpha s} +
{\Gamma(m-s+\alpha)\over \Gamma(\alpha+1)}\, d_{\alpha s} =
{\Gamma(m-s+\alpha)\over \Gamma(\alpha)}\, c_{\alpha-1,s}\,.
 \]
This, with (\ref{(6.16)}) and (\ref{(6.32)}), implies
\[
 {\Gamma(m-s+\alpha+1)\over \Gamma(\alpha+1)} [c_{\alpha s}+
\psi(m-s+\alpha+1) d_{\alpha s}]
={\Gamma(m-s+\alpha)\over \Gamma(\alpha)} [c_{\alpha-1,s} +
\psi(m-s+\alpha) d_{\alpha-1,s}], \]
establishing consistency of (\ref{(6.26)}).

The form of the recursion relations analogous to (\ref{(6.8)}) is 
(for $\alpha\ge 1$) 
 \begin{equation}
 c_{\alpha-1,s} = {1\over \alpha} (m-s+\alpha) c_{\alpha s} 
 \label{(6.33)} \end{equation} \nobreak
 if  $s-m$  is even or negative;
\begin{equation}
 c_{\alpha-1,s} = {1\over \alpha} (m-s+\alpha) c_{\alpha s}
  + {1\over\alpha} d_{\alpha s}\,,
 \label{(6.34)} \end{equation}
\nobreak \begin{equation}
d_{\alpha-1,s} = {1\over \alpha} (m-s+\alpha) d_{\alpha s}
 \label{(6.35)} \end{equation}
if $ s-m$ is  odd and positive.
Note the consistency with the remarks made in connection with (\ref{(6.28)}):
  If
$s-m$ is odd and positive, then $c_{\alpha s}$ for a value of 
$\alpha$ satisfying (\ref{(6.28)}) will generally be nonzero and 
proportional to $d_{\alpha s}$ for values of $\alpha$ violating 
(\ref{(6.28)}). 
  Otherwise, $c_{\alpha s}$ and
$d_{\alpha s}$ will vanish when (\ref{(6.28)}) is satisfied, 
because the factor $(m-s+\alpha)$ in (\ref{(6.33)}) or 
(\ref{(6.35)}) will have vanished for some larger value of 
$\alpha$. 

Let us express all the coefficients in terms of $c_{ss}$ and $d_{ss}\,$:
In analogy to (\ref{(6.9)}), we have
\begin{equation}
 e_s = {\Gamma(m+1)\over \Gamma(s+1)}\, c_{ss} 
 \label{(6.36)} \end{equation} \nobreak
 if $s-m$ is even or negative;
\begin{equation}
 e_s = {\Gamma(m+1)\over \Gamma(s+1)} [c_{ss} + \psi(m+1)
d_{ss}],
 \label{(6.37)}\end{equation} \nobreak
\begin{equation}
 f_s = -{\Gamma(m+1)\over \Gamma(s+1)} \, d_{ss} 
 \label{(6.38)} \end{equation} 
if $s-m$ is odd and positive.
The analogues of (\ref{(6.10)}) are
\begin{equation}
 c_{\alpha s} = {\Gamma(\alpha+1) \Gamma(m+1)\over
  \Gamma(m-s + \alpha+1)\Gamma(s+1)}\, c_{ss} 
 \label{(6.39)} \end{equation} \nobreak
 if $s-m$ is even or negative;
and in the contrary case,
\begin{equation}
 c_{\alpha s} = {\Gamma(\alpha+1) \Gamma(m+1)\over
\Gamma(m-s+\alpha+1) \Gamma(s+1)} \{c_{ss}
  + [\psi(m+1) - \psi(m-s+\alpha +1)] d_{ss}\},
 \label{(6.40)}\end{equation} \nobreak
\begin{equation}
 d_{\alpha s} = {\Gamma(\alpha+1) \Gamma(m+1)\over 
 \Gamma(m-s + \alpha+1)\Gamma(s+1)} \,d_{ss}\,.
 \label{(6.41)} \end{equation}
If $m-s + \alpha \ge 0$, then (\ref{(6.40)}) is unambiguous,
  and the difference of $\psi$ functions is calculable from 
 (\ref{(6.16)}) or (\ref{(6.17)}).
  But when (\ref{(6.28)}) holds, extra work is needed to obtain a 
  usable formula. 
 In that case iteration of (\ref{(6.34)}) leads to 
 \begin{equation}
 c_{\alpha s} = (-1)^{s-m-\alpha-1} {\Gamma(s-m-\alpha) 
 \Gamma(\alpha+1)\Gamma(m+1)\over \Gamma(s+1)} \,d_{ss}
\label{(6.42)} \end{equation}
  for $ \alpha <s-m$, $s-m$ odd.
This formula also can be obtained from (\ref{(6.40)}) by comparing 
residues of $\Gamma$ and $\psi$: 
 \begin{equation}
 {\psi(\varepsilon-n)\over \Gamma(\varepsilon-n)} 
 = {\Gamma'(\varepsilon-n)\over \Gamma(\varepsilon-n)^2}
= (-1)^{n-1} n! + O(\varepsilon) \qquad (n=0,1,\ldots)
 \label{(6.43)} \end{equation}
(see (\ref{(6.15)})).

\paragraph*{\bf 4.}
Finally, we investigate the relation between $T(t)$ and the
$\lambda$-means. 
 For use in (\ref{(5.3)}) we find
\begin{eqnarray}
\partial_\lambda(e^{-t\lambda^{1/2}}) &=&
e^{-t\lambda^{1/2}} \textstyle\left(-{1\over 2} t 
 \lambda^{-1/2}\right),
 \nonumber\\*
\partial^2_\lambda(e^{-t\lambda^{1/2}}) &=& e^{-t\lambda^{1/2}}
\textstyle\left({1\over 4} t^2\lambda^{-1} + {1\over 4} 
t\lambda^{-3/2}\right),
 \nonumber\\*
\partial^3_\lambda(e^{-t\lambda^{1/2}}) &=& e^{-t\lambda^{1/2}} 
\textstyle\left(-{1\over 8} t^3 \lambda^{3/2} 
 - {3\over 8} t^2\lambda^{-2}  - {3\over 8}t
\lambda^{-5/2}\right),
 \nonumber\\*
\partial^4_\lambda(e^{-t\lambda^{1/2}}) &=& e^{-t\lambda^{1/2}} 
{\textstyle\left({1\over 16} t^4 \lambda^{-2} 
 + {3\over 8} t^3 \lambda^{-5/2} + {15\over 16}
t^2\lambda^{-3} + {15\over 16} t\lambda^{-7/2}\right)},
 \label{(6.44)} \end{eqnarray}
and in general the form
\begin{equation}
 \partial^j_\lambda(e^{-t\lambda^{1/2}}) =e^{-t\lambda^{1/2}} 
\sum^j_{i=1} y_i t^i \lambda^{-j+{i\over 2}}. 
 \label{(6.45)} \end{equation}
Note that (\ref{(5.6)}) is satisfied. 
 As in the other cases, there is only one relevant term in 
(\ref{(5.3)}): 
 \begin{eqnarray}
T(t) &=& (-1)^{\alpha-1} {1\over \alpha!} \int^\infty_0
\partial^{\alpha+1}_\lambda(e^{-t\lambda^{1/2}}) \lambda^\alpha
R^\alpha_\lambda \mu \, d\lambda
 \nonumber\\*
&=& (-1)^{\alpha-1} {1\over \alpha!} \int^\infty_0 
\partial^{\alpha+1}_\lambda
(e^{-t\lambda^{1/2}}) \left[ \sum^\alpha_{s=0} a_{\alpha s}
\lambda^{{m\over 2}-{s\over 2}+\alpha} + O(\lambda^{{m\over 2}+
{\alpha\over 2}-{1\over 2}})\right]d\lambda.
 \label{(6.46)} \end{eqnarray}
However, unlike the other three cases, this time the unknown
contribution from small $\lambda$ is $O(t)$ --- 
 its order does not increase with $\alpha$. 
 Furthermore, the integrals of the individual terms in 
 (\ref{(6.46)}) do not converge at the lower limit, 
 although the integral as a whole is convergent. 
 Consequently, we must deal with integrals of the form
\begin{equation}
 \int^\infty_{\lambda_0} e^{-t\lambda^{1/2}} \lambda^{-p} d\lambda 
= 2 \int^\infty_{\omega_0} e^{-t\omega} \omega^{-2p+1}d\omega. 
 \label{(6.47)} \end{equation}
If $p<1$, we may set $\omega_0=0$ and use (\ref{(6.3)});
  if $p\ge 1$, we may use
\begin{equation}
 \int^\infty_{\omega_0} e^{-t\omega} \omega^{-n} d\omega =
\omega^{1-n}_0 \int^\infty_1 e^{-\omega_0 tu} u^{-n}\, du,
 \label{(6.48)}\end{equation}
 \begin{equation}
\int^\infty_1 e^{-t\omega} \omega^{-n}\, d\omega \equiv E_n(t),
 \label{(6.49)} \end{equation}
\begin{mathletters}\label{(6.50)}\begin{eqnarray}
 E_1(t) &=& -\gamma - \ln t - \sum^\infty_{j=1}
 {(-1)^j\over jj!}\, t^j,
 \label{(6.50a)} \\*
E_n(t) &=& {(-t)^{n-1}\over (n-1)!} [-\ln t  + \psi(n)] 
 -\sum^\infty_{\scriptstyle j=0 \atop \scriptstyle j\ne n-1}
  {(-1)^j\over (j-n+1)j!} \,t^j.
 \label{(6.50b)} \end{eqnarray}  \end{mathletters}
It can be seen that the integrals give rise to terms of the form 
(\ref{(6.24)}),
but without the restriction that $s-m$ be odd in the logarithmic terms.
Moreover, the contribution from $\lambda<1$ is analytic, hence can be
expanded as a series of positive integral powers of $t$ with unknown
coefficients. 
 From (\ref{(4.19)}) and (\ref{(6.25)})--(\ref{(6.26)}), 
 we would expect the coefficient
of $t^{-m+s}$ to be determinable if $s-m$ is even but not if $s-m$ is odd.
Therefore, we anticipate conspiracies among the coefficients in 
 (\ref{(6.45)})
which will eliminate both the logarithms and the undeterminable
coefficients when $s-m$ is even.
  Let us verify this for the case $m=1$, $\alpha=3$:
\begin{eqnarray*}
 T(t) &=& {1\over 6} \int^\infty_0 e^{-t\lambda^{1/2}}
\left({1\over 16} t^4\lambda^{-2}  + {3\over 8} t^3 \lambda^{-5/2} +
{15\over 16} t^2\lambda^{-3} + {15\over 16} t \lambda^{-7/2}\right)
 \\*
&&{} \times \bigl[a_{30} \lambda^{7/2} + a_{31}\lambda^3 + a_{32}
\lambda^{5/2} + a_{33}\lambda^2 + O(\lambda^{3/2})\bigr]\,d\lambda
 \\
&=& {1\over 48} \int^\infty_0 e^{-t\omega} \big(a_{30} t^4 \omega^4 + 6a_{30}
t^3\omega^3 + a_{31} t^4\omega^3
 + 15 a_{30} t^2\omega^2+ 6a_{31} t^3 \omega^2 + a_{32}t^4
\omega^2
 \\*
&&{} +15 a_{30} t\omega + 15 a_{31} t^2\omega + 6a_{32} t^3\omega +
a_{33} t^4\omega
 + 15a_{31} t + 15 a_{32} t^2 + 6a_{33}t^3
 \\*
&&{} + 15 a_{32} t\omega^{-1} + 15 a_{33}t^2 \omega^{-1} + 15 a_{33}
t\omega^{-2} + O(t\omega^{-3})\big)\,d\omega
 \\
&=& {1\over 48} \bigg\{105 a_{30} t^{-1} 
 + 48 a_{31} + 23 a_{32}t + 7a_{33}t^2 
\\*
&& {} + \int^{\omega_0}_0 e^{-t\omega} (15 t^2\omega^{-6})
  [O(\omega^6) - a_{32}\omega^5] \omega\, d\omega
 + \int^{\omega_0}_0 e^{-t\omega} (15t\omega^{-7}) O(\omega^6)
\omega\, d\omega + O(t^3)
 \\*
&&{}+ 15 a_{32} tE_1(\omega_0t) + 15a_{33} t^2 E_1(\omega_0t) 
 + 15a_{33} \omega^{-1}_0 tE_2(\omega_0t)
\left. +\int^\infty_{\omega_0} e^{-t\omega} O(t\omega^{-s})
d\omega\right\}.
  \end{eqnarray*}
(Here the first four terms come from integrating all terms in the 
middle member involving $\omega$ to a nonnegative power. 
 To represent accurately the integral over small $\omega$ of the 
remaining terms, it has been necessary to go back to the first 
member and to note that the quantity in square brackets there is 
$O(\omega^6)$ by virtue of (\ref{(6.46)}) and (\ref{(5.5)}). 
  Of the terms involving $t\lambda^{-7/2}$, the first two have 
already been accounted for in the explicit integrations, and the 
remainder of the square bracket is still $O(\omega^6)$ as
$\omega\downarrow 0$; 
 of the terms involving $t^2\lambda^{-3}$, the first three have been 
accounted for, and the subtraction of the third of these needs to 
be represented explicitly in our formula; 
 the terms associated with the rest of
$\partial^4_\lambda(e^{-t\lambda^{1/2}})$ are $O(t^3)$.) 
 Substituting from
(\ref{(6.48)})--(\ref{(6.50)}),
  and noting that the two quantities represented by
``$O(\omega^6)$'' in the foregoing expression are the {\em same}, 
 we obtain
\begin{eqnarray*}
T(t) &=& {1\over 48} 
\bigg\{105 a_{30}t^{-1} + 48a_{31} + 23 a_{32}t +
7a_{33} t^2- 15t^2a_{32}\omega_0
 \\*
&&{} + \int^{\omega_0}_0 15 t\,O(\omega^0) 
\bigl[\omega t +
O(t^2) + 1-\omega t\bigr]\,d\omega
+ O(t^3) + tO(\omega^{-2}_0 E_3(\omega_0t)) 
 \\*
&&{} + 15 a_{32}t [-\gamma - \ln \omega_0t + \omega_0t + O(t^2)]
 + 15 a_{33}t^2 [-\gamma-\ln \omega_0t + O(t)]
 \\*
&&{} + 15 a_{33}\omega^{-1}_0 t [-\omega_0t (-\ln \omega_0 t 
 + \psi(2)) + 1 + O(t^2)]\bigg\}
\\
&=& {1\over 48} \bigg\{105 a_{30} t^{-1} + 48a_{31}
 - 15 a_{32} t\ln \omega_0 t
\\*
&&{} + t\biggl[23 a_{32} + 15\int^{\omega_0}_0 O(\omega^0) d\omega 
- 15 \gamma
a_{32} + 15 a_{33} \omega^{-1}_0\biggr]
 \\*
&&{} + t^2[7a_{33} - 15 a_{32} \omega_0 + 15 a_{32}\omega_0
  - 15 \gamma a_{33}
 + 15 \gamma a_{33} - 15 a_{33}]
 \\*
 &&{} + O(t^3)
 + \omega^{-2}_0 t\left[O\left({\textstyle{1\over 2}}
-\omega_0t\right) +
O(t^2\ln t)\right]\bigg\}
 \\
&=& {1\over 48} \big\{105 a_{30}t^{-1} + 48 a_{31} 
 - 15 a_{32} t \ln \omega_0 t
 + \text{const.} \times t - 8a_{33} t^2 
 \\*
 &&{}+ O(t^3\ln t)
 + O(\omega^{-2}_0t) + O(\omega^{-1}_0t^2)\big\}.
  \end{eqnarray*}
The bothersome term at the end could be removed by taking $\omega_0 
\to \infty$. 
 Now this term and its predecessor can be traced to the first two 
terms in a Taylor expansion of $e^{-t\omega}$ in (\ref{(6.49)}) with 
$n=3$. 
 In analogy to the argument leading from (\ref{(4.16)}) to 
(\ref{(4.18)}), 
  we may transfer this part of the integral from the high-$\omega$ 
  account to the low-$\omega$ account; 
 this simply amounts to adding something to the ``$O(\omega^6)$'' 
term, and the same cancellation of $O(t^2)$ terms just observed for 
that term will occur for this other contribution as well. 
 Thus we have derived the expected form, 
 \[
 T(t) = {\textstyle{35\over 16}} a_{30} t^{-1} + a_{31}
  - {\textstyle{5\over 16}} a_{32} t\ln t + e_2 t 
 - {\textstyle{1\over 6}} a_{33} t^2 +O(t^3\ln t),
 \]
where $e_2$ is undeterminable from the asymptotic expansion of
$R^3_\lambda\mu$. 
 Furthermore, the coefficients agree with those calculated
from (\ref{(6.24)})--(\ref{(6.27)}) and 
 (\ref{(4.19)})--(\ref{(4.21)}).

\paragraph*{\bf Summary:\/}
  The coefficients $b_s$ in the asymptotic expansion of the heat 
  kernel, $K(t)$, are in one-to-one correspondence with the 
  ``diagonal'' coefficients $a_{ss}$ in the asymptotic expansions 
  of the Riesz means with respect to $\lambda$, $R^s_\lambda\mu$. 
 The nondiagonal coefficients, $a_{\alpha s}$, in the expansions of 
the $R^\alpha_\lambda \mu$ differ only by numerical factors from 
the $a_{ss}$; 
 because these factors may vanish, it is not possible to express 
$a_{ss}$ in terms of $a_{\alpha s}$ if $\alpha$ is too small. 
 Similarly, the coefficients $e_s$ and $f_s$ in the expansion of the 
cylinder kernel, $T(t)$, are in one-to-one correspondence with 
the diagonal coefficients $c_{ss}$ and $d_{ss}$ in the 
expansions of the Riesz means with respect to $\omega$, 
$R^s_\omega\mu$. 
 The connection in this case involves a two-termed equation, 
(\ref{(6.37)}). 
  Again the $c_{\alpha s}$ and $d_{\alpha s}$ can be expressed in 
  terms of the $c_{ss}$ and $d_{ss}$, but not conversely if 
  $\alpha$ is too small. 
 Finally, and perhaps most significantly, the $c_{ss}$ and $d_{ss}$ 
(or the $e_s$ and $f_s$) contain information which is not contained 
in the $a_{ss}$ (or the $b_s$) (but not conversely). 
 The $c_{ss}$ for $s-m$ odd and positive are ``new spectral 
invariants'' independent of the $a_{ss}\,$. In the case of the heat 
and cylinder kernels, the $c_{ss}$ contain nonlocal geometrical 
information, while the $a_{ss}$ are strictly local.

 \section{Conclusion}\label{sec7}

 Motivated by applications in spectral asymptotics and quantum 
field theory, we have investigated Riesz means in this setting:
 Two functions, $K(t)$ and $T(t)$, are related to a function 
$\mu(\lambda)$ by the Stieltjes integrals
 \begin{equation}
 K(t) =\int_0^\infty e^{-\lambda t} \, d\mu,
 \qquad T(t) =\int_0^\infty e^{-\omega t} \, d\mu,
 \label{(7.1)} \end{equation}
 where $\omega = \sqrt \lambda$, and they have the asymptotic 
expansions
\begin{equation}
 K(t) \sim \sum^\infty_{s=0} b_s t^{-{m\over 2}+{s\over 2}} ,
 \qquad
T(t) \sim \sum^\infty_{s=0} e_s t^{-m+s} + 
 \sum^\infty_{\scriptstyle s=m+1\atop
\scriptstyle s-m\text{ odd}} f_s t^{-m+s} \ln t,
 \label{(7.2)} \end{equation}
as $t\downarrow0$, for some $m\in \ZZ^+$.
 The behavior of $\mu$ as $\lambda \to +\infty$ is characterized by 
the constants $a_{ss}\,$, $c_{ss}\,$, $d_{ss}\,$, where
 \begin{equation}
 R^\alpha_\lambda\mu(\lambda) \sim
  \sum^\alpha_{s=0} a_{\alpha s} \lambda^{{m
\over 2}-{s\over 2}}, 
 \qquad
R^\alpha_\omega\mu \sim \sum^\alpha_{s=0} c_{\alpha s}
\omega^{m-s} + \sum^\alpha_{\scriptstyle s=m+1\atop 
 \scriptstyle s-m\text{ odd}} d_{\alpha s} \omega^{m-s} \ln \omega,
 \label{(7.3)} \end{equation}
 $R^\alpha_\lambda\mu$ and $R^\alpha_\omega\mu$ being the Riesz 
means with respect to $\lambda$ and $\omega$, defined by, for 
example,
 \begin{mathletters}\label{(7.4)}\begin{eqnarray}
 R^\alpha_\omega\mu(\omega^2) &=&
 \int_{\tau=0}^\omega \left(1 - {\tau\over\omega}\right)^\alpha 
 \, d\mu(\tau^2) 
 \label{(7.4a)}  \\*
 &=& \alpha!\, \omega^{-\alpha} \int_0^\omega d\tau_1 \cdots
 \int_0^{\tau_{\alpha-1}} d\tau_\alpha \,
 \mu(\tau_\alpha\!^2).  
 \label{(7.4b)} \end{eqnarray} \end{mathletters}

 We have shown that  $a_{ss}$ and $b_s$ contain the same 
information, being related by (\ref{(6.9)}),
  and that $(c_{ss}, d_{ss})$ 
  and $(e_s, f_s)$ contain the same information, being related by
(\ref{(6.36)})--(\ref{(6.38)}).
 The inverses of the cited formulas are
 \begin{equation}
 a_{ss} = {\Gamma(s+1) \over 
 \Gamma\bigl({m\over2} + {s\over2} +1\bigr)}\, b_s \,; 
 \label{(7.5)} \end{equation}
 \begin{equation}
  d_{ss} = -\, {\Gamma(s+1) \over  \Gamma(m+1)} \,f_s \quad
 \text{if $s-m$ is odd and positive}; 
 \label{(7.6)} \end{equation}
 \begin{equation}
 c_{ss} = {\Gamma(s+1) \over  \Gamma(m+1)}
 [e_s + \psi(m+1) f_s] \quad
  \text{if $s-m$ is odd and positive}, 
 \label{(7.7)} \end{equation}
 \begin{equation}c_{ss} = {\Gamma(s+1) \over  \Gamma(m+1)} \, e_s 
\quad  \text{if $s-m$ is even or negative}.
  \label{(7.8)} \end{equation}

 Finally, these two collections of quantities are related to each 
other, but in an asymmetrical way.
 Specializing (\ref{(4.19)})--(\ref{(4.21)}) and 
 (\ref{(4.25)})--(\ref{(4.26)}), we have
\begin{equation}
 c_{s s} \text{ is undetermined by $a_{ss}$ if $s-m$ is
odd and positive};
 \label{(7.9)} \end{equation}
\begin{equation}
 d_{s s} = (-1)^{m+1} {1\over (s-m-1)!\, m!}\,
{\Gamma\big({s\over 2}-{m\over 2}\big)\over \Gamma\big(- {m\over
2}-{s\over2}\big)} \,a_{s s} 
 \quad
\text{if  $s-m$ is odd and positive};
 \label{(7.10)}\end{equation}
\begin{eqnarray}
 c_{s s} = \left[2^{-s}+\sum^{s-1}_{\scriptstyle j=0 \atop 
 \scriptstyle j\text{ even}}(-1)^{s-j}
{(m-s+j+1)^{-1}\over j!\,(s-1-j)!}\, {\Gamma\big({j\over 2}
  + {1\over 2}\big)\over \Gamma\big({j\over 2} 
 +{1\over 2}-s\big)}\right] a_{s s}& 
\nonumber\\*
 \text{if  $s-m$ is even or negative};&
 \label{(7.11)}  \end{eqnarray}
\begin{eqnarray}
 a_{s s} = \left[-{1\over 2} \sum^{s-1}_{j=\big[{s\over
2}\big]} (-1)^{s-j} {\big({m\over 2}-{s\over 2} + j+1)^{-2} \over
j!\,(s-1-j)!}\, {\Gamma(2j+2)\over \Gamma(2j+2-s)}\right] 
 d_{ss}&
 \nonumber\\*
 \text{if  $s-m$ is odd and positive};&
 \label{(7.12)} \end{eqnarray}
\begin{eqnarray}
 a_{s s} = \left[2^s +\sum^{s-1}_{j=\big[{s\over 2}\big]}
(-1)^{s-j} {\big({m\over 2}-{s\over 2} +j+1\big)^{-1} \over
j!\,(s-1-j)!}\, {\Gamma(2j+2)\over \Gamma(2j+2-s)}\right]\, c_{ss}&
\nonumber\\*
 \text{if  $s-m$ is even or negative}.&
\label{(7.13)} \end{eqnarray} 
These relations obviously induce similar relations between $b_s$ 
and $(e_s, f_s)$.
 The quantities $c_{ss}$ ($s-m$ odd and positive)
 summarize the information contained in the asymptotics of $T$ but 
not in that of $K$.

In the context of spectral theory of differential operators,
 the quantities $b_s\,$, $e_s\,$, $f_s$ are more accessible to 
calculation, because $K(t)$ and $T(t)$ are Green functions or their 
traces,
 but $a_{ss}\,$, $c_{ss}\,$, $d_{ss}$ are more fundamental,
 since $\mu(\lambda)$ represents a spectral decomposition
 (the integral kernel of the spectral projections in the local 
case, the density of eigenvalues in the traced case).
 We expect that further work will bring the technically more 
difficult case of the Wightman function (and other Green functions 
of wave equations) into the same picture.

 This picture clarifies and unifies results that have appeared in 
the literature piecemeal or imprecisely.
 One sees that the small-$t$ expansion of $K$ is not related in a 
one-to-one way with the formal large-$\lambda$ expansion of~$\mu$
 (i.e., the $a_{0s}$);
 rather, the heat coefficients $b_s$ can involve constants of 
integration from the Riesz averaging of~$\mu$ over small~$\lambda$.
 This explains how the formal expansion of~$\mu$ (and the 
associated zeta-function poles) can be strikingly different in even 
and odd dimensions, while the heat-kernel expansion is notoriously 
dimension-independent.
 (See Remark below (\ref{(6.10)}).)
 Further constants of integration appear in the passage from 
$\lambda$-means to $\omega$-means;
 these carry nonlocal geometrical information in the spectral 
context.
 They are intimately related to the potential appearance of 
logarithmic terms 
 $(f_s, d_{ss})$.
On the other hand, the numerical relationships among the various 
series are independent of the spectral application;
 in particular, they will be universal for all (positive, elliptic) 
operators of a given order and dimension.
 (Only order~2 has been treated here.)

 To conclude, we apply some of these formulas to the simple 
examples studied in Sec.~\ref{sec2}
  and compare the results to the known 
spectral densities for those cases.
 In these examples, $m=1$.

\subsection*{Case ${\cal M}=\RR$}

Formula (\ref{(2.7)}) for the heat kernel states that

 \begin{mathletters}\label{(7.14)} \begin{equation}
  b_s = (4\pi)^{-1/2} \delta_{s0} \quad \text{if $y=x$}, 
 \label{(7.14a)} \end{equation}
 \begin{equation} 
 b_s = 0 \quad\text{for all $s$ \quad if $y \ne x$}.
 \label{(7.14b)} \end{equation} \end{mathletters}
 From (\ref{(7.5)}), then, one has for the Riesz means of the 
spectral function 
\begin{mathletters}\label{(7.15)} \begin{equation}
  a_{ss} = {1\over\pi}\, \delta_{s0} \quad  \text{if $y=x$}, 
 \label{(7.15a)} \end{equation}
 \begin{equation}a_{ss}=0\quad\text{for all $s$ \quad if $y \ne x$}.
 \label{(7.15b)} \end{equation}\end{mathletters}
The eigenfunction expansion in this case is the Fourier transform, 
so the exact spectral density is 
 \begin{equation}
dE_\lambda(x,y) 
= {1\over 2\pi} \sum_{\mathop{\rm sgn} k} e^{ik(x-y)} \,d|k| 
 = {1\over \pi} \, \cos[\omega(x-y)] \,d\omega
 \qquad(\sqrt\lambda = \omega = |k|). 
 \label{(7.16)} \end{equation}
 Thus
\begin{mathletters}\label{(7.17)} \begin{equation}
E_\lambda(x,x) = {\omega \over \pi} = {\sqrt\lambda \over \pi}\,, 
 \label{(7.17a)} \end{equation}
 \begin{equation}E_\lambda(x,y) = {1\over\pi} \, 
 {\sin[\omega(x-y)]\over x-y} \quad \text{if $y\ne x$}. 
 \label{(7.17b)} \end{equation}\end{mathletters}
 (Constants of integration are fixed by (\ref{(3.1)}).)
 Clearly (\ref{(7.17a)}) is consistent with (\ref{(7.15a)}).
 Equation (\ref{(7.15b)}) indicates that (\ref{(7.17b)})
  is ``distributionally small''\negthinspace,
 or decays rapidly at infinity in the Ces\`aro 
 sense.\cite{EK,E96,EF}
 To see this by elementary, classical methods, one would use
 (\ref{(3.9)}),
 \begin{equation}
 R_\lambda^s E_\lambda(x,y) = 
 {1\over\pi} \int_0^\lambda \left(1 - {\sigma\over\lambda} 
\right)^s 
 \, d\left\{ {\sin[\sqrt\sigma(x-y)] \over x-y} \right\}, 
 \label{(7.18)} \end{equation}
and verify that this object is 
 $o\bigl(\lambda^{1-s \over 2}\bigr) $
as $\lambda \to \infty$.
 This can be done by repeated integration by parts
 (most easily after changing variable from $\sigma$ to
 $\tau =\sqrt\sigma$).

 For the less familiar cylinder kernel, (\ref{(2.9)}) gives
 \begin{equation}
 f_s =0, 
 \label{(7.19)} \end{equation}
 \begin{mathletters}\label{(7.20)}\begin{equation}
 e_s = {1\over \pi} \, \delta_{s0} \quad
 \text{if $y=x$}, 
 \label{(7.20a)} \end{equation}
 and if $y \ne x$,
 \begin{equation}
  e_s = \cases{ 
 0 &\text{if $s=0$ or $s$ is odd,}\cr
 \noalign{\smallskip}
  \displaystyle 
 (-1)^{{s\over 2} +1} \,{1\over \pi}\, {1\over (x-y)^s}
 &\text{if $s$ is even and positive.} }
 \label{(7.20b)} \end{equation}\end{mathletters}
  From (\ref{(7.6)})--(\ref{(7.8)}), 
 this corresponds to the $\omega$-means
 \begin{equation}
 d_{ss} =0, 
 \label{(7.21)} \end{equation}
 \begin{mathletters}\label{(7.22)}\begin{equation}
 c_{ss} = {1\over\pi} \, \delta_{s0} \quad
 \text{if $y=x$}, 
 \label{7.22a)} \end{equation}
 and if $y \ne x$,
 \begin{equation} 
 c_{ss} = \cases{ 
 0 &\text{if $s=0$ or $s$ is odd,}\cr
 \noalign{\smallskip} 
 \displaystyle
 (-1)^{{s\over 2} +1} \,{s!\over \pi}\, {1\over (x-y)^s}
 &\text{if $s$ is even and positive.} }
 \label{(7.22b)} \end{equation}\end{mathletters}
 One observes that (\ref{(7.15)}) and (\ref{(7.21)})--(\ref{(7.22)})
  are related precisely as prescribed in 
 (\ref{(7.9)})--(\ref{(7.13)}).
 Again, (\ref{(7.21)})--(\ref{(7.22)})
  could be established directly from (\ref{(7.4a)}) and 
(\ref{(7.17)}) by integration by parts.
 (The earnest student who actually attempts these tedious exercises 
will find that certain endpoint terms that vanished in the previous 
case will produce the nonvanishing $c_{ss}$ in this case.)
 The present case, however, is more easily treated by 
(\ref{(7.4b)}).

\subsection*{Case ${\cal M} = \RR^+$}

 Let us consider only the Dirichlet boundary condition.
 The heat-kernel asymptotics, and hence the $\lambda$-means 
$a_{ss}\,$, are again trivial.
 The effect of the boundary is seen (as dictated by (\ref{(7.9)}))
 only in the even-order coefficients of the cylinder kernel and the 
associated $\omega$-means: 
 from (\ref{(2.12)}), for $s>0$,
 \begin{mathletters}\label{(7.23)}\begin{equation}
 e_s = {(-1)^{s/2} \over \pi (2x)^s} = {1\over s!}\, c_{ss} \quad
 \text{if $y=x$ and $s$ is even}, 
 \label{(7.23a)} \end{equation}
 \begin{equation}
 e_s = -\, {(-1)^{s/2} \over \pi} 
 \left[ {1\over (x-y)^s} - {1\over (x+y)^s} \right]
 = {1\over s!}\, c_{ss} \quad
 \text{if $y\ne x$ and $s$ is even}, 
 \label{(7.23b)} \end{equation} \end{mathletters}
\begin{equation}
 e_s = 0 = c_{ss} \quad\text{if $s$ is odd}. 
 \label{(7.24)} \end{equation}

 The eigenfunction expansion in this case is the Fourier sine 
transform, so
 \begin{equation}
 dE_\lambda(x,y) = {2\over\pi} \, \sin(\omega x) \sin(\omega y)
 \, d\omega. 
  \label{(7.25)} \end{equation}
 Let us check consistency only for the diagonal values.
 We have
 \begin{equation}
  E_\lambda(x,x) = 
 {\omega\over \pi} -  {\sin{(2\omega x)} \over 2\pi x} \,. 
  \label{(7.26)} \end{equation}
 The first term is the same as (\ref{(7.17a)}) and need not be 
discussed further. 
 The $\lambda$-means  of the second term are shown to vanish 
exactly as for (\ref{(7.18)}).
 The $\omega$-means of the second term can be calculated by 
 (\ref{(7.4b)}) 
exactly as for (\ref{(7.22b)}),
  and they reproduce (\ref{(7.23a)}).

\subsection*{Case ${\cal M}=S^1$}

The eigenfunction expansion is the full Fourier series, so
 \begin{eqnarray}
dE_\lambda(x,y) &=& {1\over 2L} \sum_{\mathop{\rm sgn} k} 
 e^{ik(x-y)} \sum_{n=0}^\infty \delta \left(|k| - {n\pi\over L} 
\right) \, d|k| 
 \nonumber\\* 
 &=& \left[ {1\over 2L} \, \delta(\omega)
 + {1\over L} \cos{[\omega(x-y)]} \sum_{n=1}^\infty 
 \delta \left(\omega - {n\pi \over L} \right) \right]\, d\omega.
\label{(7.27)} \end{eqnarray}
 The traced expansions of the Green functions correspond to the 
eigenvalue distribution function,
 \begin{equation}
  dN(\lambda) = \int_{-L}^L dE_\lambda(x,x)\, dx 
 = \left[\delta(\omega) + 2 \sum_{n=1}^\infty \delta \left(
\omega - {n\pi\over L} \right) \right] \, d\omega. 
 \label{(7.28)} \end{equation}
The calculation of Riesz means leads to sums that can be regarded 
as trapezoidal-rule approximations to integrals of the types 
already considered.
 Therefore, the new features of this case can be sought in the 
Euler--Maclaurin formula (Ref.~\onlinecite{EK}, Sec.~1.8)
 for the difference between the sum and the integral. 
 The terms in that formula involve only the odd-order derivatives 
of the integrand at the endpoints of the interval.
 In an integral such as (\ref{(7.4a)}) or (\ref{(7.18)}),
 the derivatives of the factor such as $(1 - \tau/\omega)^s$ to all 
relevant orders will vanish at the upper limit, so only the lower 
endpoint can contribute. 
 In the $\lambda$-means (\ref{(7.18)}), where this Riesz factor is
 $(1-\tau^2/\lambda)^s$,
 the odd-order derivatives all vanish at $0$ by  virtue of factors 
of $\tau$ or $\sin[\tau(x-y)]$.
 This is necessary for consistency with the trivial heat-kernel 
expansion, (\ref{(2.7)}) and (\ref{(2.17)}).
For the $\omega$-means, we need to calculate
 \begin{equation}
 {1\over L} \left({\pi\over L}\right)^p {d^p \over d\tau^p}
 \left[ \left( 1 - {\tau\over\omega}\right)^s 
 \cos{[\tau(x-y)]} \right].  
 \label{(7.29)} \end{equation}
 When $y=x$, this derivative is of order $\omega^{-p}$;
 in fact, it equals
 \begin{equation}
  {(-1)^p \pi^p s! \over L^{p+1} \omega^p (s-p)!} \,. 
\label{(7.30)} \end{equation}
 (In the off-diagonal case, the Euler--Maclaurin series does not
 yield an expansion of the desired type --- at least, not without a 
resummation, which we shall not attempt here.)
To read off $c_{ss}$ (for $s\ge 2$) we need to look at $p=s-1$
 (see (\ref{(7.3)})) and, according to Euler--Maclaurin, multiply by
 $(-1)^p B_s/s!$,
 where $B_s$ is a Bernoulli number.
 Thus we arrive at
 \begin{equation} 
 c_{ss} = {\pi^{s-1} B_s \over L^s}\,, \qquad
 e_s = {\pi^{s-1} B_s \over L^s s!}\,.   
 \label{(7.31)} \end{equation}
 This is, in fact, the same as (\ref{(2.14)})--(\ref{(2.15)})
  with $y=x$, because
 \begin{equation}
 {1\over 2}\,{\sinh z \over \cosh z -1} = \coth {z\over 2}
 = {1\over z} \left[ 1 + \sum_{s=2}^\infty {B_s \over s!} \, z^s 
\right],
  \label{(7.32)} \end{equation}
 \begin{equation} 
 B_s = 0 \quad\text{if $s$ is odd and $s>2$}. 
 \label{(7.33)} \end{equation}

 \subsection*{Case ${\cal M}=I$}

 For Dirichlet boundary conditions the eigenfunction expansion is 
the Fourier sine series, so
 \begin{equation}
 dE_\lambda(x,y) = {2\over L} \,\sin{(\omega x)}\sin{(\omega y)}
 \sum_{n=1}^\infty \delta\left(\omega - {n\pi\over L}\right)\, 
d\omega 
 \label{(7.34)} \end{equation}
 and
 \begin{equation}
 dN(\lambda ) = \sum_{n=1}^\infty \delta \left(\omega-{n\pi \over 
L}\right) \, d\omega . 
 \label{(7.35)} \end{equation}
 The novel feature of this case is that the trace (\ref{(2.23)}) of 
the heat kernel contains a nontrivial term, of order $t^0$, 
 and a similar term appears in the trace (\ref{(2.24)}) of the 
cylinder kernel. 
 The corresponding Riesz means are
 \begin{equation} {\textstyle
 a_{11} = b_1 = -\, {1\over2}\,, \qquad
c_{11} = e_1 = -\, {1\over2}\,. } 
  \label{(7.36)} \end{equation}
 In an Euler--Maclaurin calculation these terms arise from the 
innocent fact that the initial term of the trapezoidal rule,
 \begin{equation}
 {\textstyle {1\over2}} \delta(\omega -0)\, d\omega, 
 \label{(7.37)}  \end{equation}
 is missing from (\ref{(7.35)}) and needs to be subtracted ``by 
hand''\negthinspace.

 The reader will have noted that in most of these instances it is 
easier to obtain the Riesz means from the Green functions via
 (\ref{(7.5)})--(\ref{(7.8)})
  and Section~\ref{sec2} than to calculate the Riesz means directly, 
even though the spectral functions are known exactly.
 (Since the direct calculations are merely consistency checks for 
us,
 we have carried out some of the calculations only far enough to 
demonstrate the matching of the first few terms, rather than 
constructing complete proofs.
 The claims in this section are to be understood in that spirit.)

\section*{Acknowledgments}

Most of this paper was written during the early 1980s when I was supported
by the National Science Foundation under Grant No.\ PHY79--15229.
 During the earliest stage I enjoyed the hospitality of the Institute for
Theoretical Physics (University of California at Santa Barbara). 
 At both
ITP and Texas A\&M University I was able to discuss the ideas in seminars
and receive helpful comments from members of the audiences, including John
Cardy, David Gurarie, and especially Robert Gustafson, who has supplied a
direct proof of the identity (\ref{(4.28)}) to which I had been led indirectly.
 Michael Taylor suggested that the ``cylinder kernel'' of a differential
operator $H$ is an object worthy of study. 
 Above all, I am grateful to
Ricardo Estrada, who reawakened my interest in this project after it lay
dormant for ten years, and advanced it in a different direction in our
joint paper.~\cite{EF}
Any integral or special-function identity cited above without a reference
probably came from the indispensable Ref.~\onlinecite{GR}.

\appendix
\section*{Proof of (\protect\ref{(4.28)}) \\*
\rm by R. A. Gustafson}  

We shall prove the following identity:
\begin{eqnarray}
1 &=& \left[\sum^{\alpha-1}_{j=\big[{\alpha\over 2}\big]}
(-1)^{\alpha-j} {\big({1\over 2}z+j+1\big)^{-1}\over j!\,(\alpha-1-j)!}
{\Gamma(2j+2)\over \Gamma(2j+2-\alpha)} +2^\alpha\right]
 \nonumber\\*
&&{} \times \left[\sum^{\alpha-1}_{\scriptstyle j=0\atop \scriptstyle 
 j\text{ even}}
(-1)^{\alpha-j} {(z+j+1)^{-1}\over j!\,(\alpha-1-j)!} {\Gamma\big({1\over
2}(j+1)\big)\over \Gamma\big({1\over 2}(j+1)-\alpha\big)} +
2^{-\alpha}\right],
 \label{(A.1)} \end{eqnarray}
where $\alpha\ge 1$ is an integer and $z\in\CC$ is not a pole of either
factor. 
 The key fact used is the ${}_3F_2$ transformation formula\cite{B},
\begin{equation}
 {}_3F_2 \left(\matrix{a,b,-n\cr &;1\cr e,f\cr}\right) = {(e-
a)_n(f-a)_n \over (e)_n(f)_n} \,{}_3F_2 \left(\matrix{1-s,a,-
n\hfill\cr &;1\cr 1+a-f-n, 1+a-e-n\hfill\cr}\right) 
 \label{(A.2)} \end{equation}
where $s=e+f-a-b+n$.

First we compute the sum in the first factor of (\ref{(A.1)}).
  With a little work
one finds
\begin{eqnarray}
 &&\sum^{\alpha-1}_{j=\big[{\alpha\over 2}\big]} (-1)^{\alpha-j} 
{\big({1\over 2}z+j+1\big)^{-1}\over j!\,(\alpha-1-j)!} 
{\Gamma(2j+2)\over \Gamma(2j+2-\alpha)} 
 \nonumber\\*
&=& {(-1) (2\alpha-1)!\over \big({z\over 2}+\alpha\big) [(\alpha-
1)!]^2} \,{}_3F_2 \left(\matrix{-{z\over 2}-\alpha, -{\alpha\over 
2} + {1\over 2}, -{\alpha\over 2}+1\hfill\cr &;1\cr -{z\over 2} - 
\alpha+1, -\alpha+ {1\over 2}\hfill\cr}\right). 
 \label{(A.3)} \end{eqnarray}

{\em Case 1: $\alpha$ is even.} 
 In (\ref{(A.2)}) set $a = -{\alpha\over 2} + {1\over 2}$, 
 $b=-{z\over 2}-\alpha$, and $-n = -{\alpha\over 2}+1$; 
 then the expression (\ref{(A.3)}) becomes 
 \begin{equation}
 {(-1)(2\alpha-1)!\over \big({z\over 2}+\alpha\big)
[(\alpha-1)!]^2} {\big(-{z\over 2} -{\alpha\over 2}+{1\over 2}\big)
_{\alpha/2-1} \big(-{\alpha\over 2}\big)_{\alpha/2-1}\over
\big(-{z\over 2}-\alpha+1\big)_{\alpha/2-1} \big(-\alpha+{1\over
2}\big)_{\alpha/2-1}} \,
 {}_3F_2\left(\matrix{1, -{\alpha\over 2}+{1\over 2},
 -{\alpha\over 2}+1\hfill\cr &;1\cr {z\over 2} + {3\over 
2},2\hfill\cr}\right). 
 \label{(A.4)} \end{equation}
Now observe that
\begin{eqnarray}
{}_3F_2 \left(\matrix{1,-{\alpha\over 2} + {1\over 2}, -{\alpha\over 
2}+1\hfill\cr &;1\cr {z\over 2}+{3\over 2},2\hfill\cr} \right) &=& 
{\big(z+  {1\over 2}\big)\over \big(-{\alpha\over2}-{1\over 2}\big)
  \big(-{\alpha\over 2}\big)} 
 \left[{}_3F_2 \left(\matrix{1, -{\alpha\over 2} - {1\over 2}, -
{\alpha\over 2}\hfill\cr  &;1\cr {z\over 2}+{1\over 
2},1\hfill\cr}\right) -1\right] 
 \nonumber\\*
&=&{\big(z+{1\over 2}\big)\over \big(-{\alpha\over2}-{1\over 2}\big)
  \big(-{\alpha \over 2}\big)}
  \left[{}_2F_1 \left(\matrix{-{\alpha\over2}-{1\over 2}, 
 -{\alpha \over 2}\hfill\cr &;1\cr {z\over 2}+{1\over 
2}\hfill\cr}\right) - 1\right]. 
 \label{(A.5)} \end{eqnarray}
Hence the expression in (\ref{(A.3)}) becomes 
\begin{eqnarray}
&&{(-1) (2\alpha-1)!\over \big({z\over 2}+\alpha\big) [(\alpha-
1)!]^2} {\big(z+{1\over 2}\big)\over 
 \big(-{\alpha\over2}-{1\over 2}\big) 
\big(-{\alpha\over 2}\big)} \left[{}_2F_1 
 \left(\matrix{-{\alpha\over2} -{1\over 2}, 
 -{\alpha\over 2}\hfill\cr &;1\cr 
 {z\over 2}+ {1\over 2}\hfill\cr}\right) -1\right] 
 \nonumber\\*
&&\quad {}\times {\big(-{z\over 2} -{\alpha\over 2}+{1\over 
2}\big)_{\alpha/2-1} \big(-{\alpha\over 2}\big)_{\alpha/2-1} \over 
\big(-{z\over 2}-\alpha+1\big)_{\alpha/2-1} \big(-\alpha+{1\over 
2}\big)_{\alpha/2-1}}\,. 
 \label{(A.6)} \end{eqnarray}
By the Gauss summation theorem for ${}_2F_1\,$, this is
\begin{eqnarray}
&& {(-1)(2\alpha-1)!\, \big(z+{1\over 2}\big) \big(-{z\over 2} -
{\alpha\over 2}+{1\over 2}\big)_{\alpha/2-1} \big(-{\alpha\over 
2}\big)_{\alpha/2-1}\over \big({z\over 2}+\alpha\big) [(\alpha-
1)!]^2 \big(-{\alpha\over2}-{1\over 2}\big) 
 \big(-{\alpha\over 2}\big) 
\big(-{z\over 2}-\alpha+1\big)_{\alpha/2-1} \big(-\alpha+{1\over 
2}\big)_{\alpha/2-1}} 
 \nonumber\\*
&&\quad{} \times \left[{\Gamma\big({z\over 2}+{1\over 2}\big) 
\Gamma\big({z\over 2} + \alpha+1\big)\over \Gamma\big({z\over 2} + 
{\alpha\over 2}+1\big) \Gamma\big({z\over 2} + {\alpha\over 2} + 
{1\over 2}\big)} -1\right]. 
 \label{(A.7)} \end{eqnarray}
After carrying out a long, easy calculation we obtain:
\begin{equation}
 \sum^{\alpha-1}_{j=0} (-1)^{\alpha-j} {\big({1\over 2}z+j-
1\big)^{-1} \over j!\,(\alpha-1-j)!} {\Gamma(2j+2) \over \Gamma(2j+2-
\alpha)}
  = 2^\alpha {\big({z\over 2}+{1\over 2}\big)_{\alpha/2} 
\over \big({z\over 2} + {\alpha \over 2} +1\big)_{\alpha/2}} -
2^\alpha, 
 \label{(A.8)} \end{equation}
recalling that $\alpha$ is assumed to be even.

{\em Case 2: $\alpha$ is odd.}
  We use (\ref{(A.3)}), now setting 
 $a = -{\alpha\over 2}+1$, $b=-{z\over 2}-\alpha$ and 
 $-n = -{\alpha \over 2}  + {1\over 2}$. 
 Doing a calculation very similar to the previous case, we obtain
\begin{equation}
 \sum^{\alpha-1}_{j=0} (-1)^{\alpha-j} {\big({1\over 2} 
z+j+1\big)^{-1} \Gamma(2j+2)\over j!\,(\alpha-1-j)! \Gamma(2j+2-
\alpha)} 
 = 2^\alpha {\big({z\over 2}+{1\over 2}\big)_{\alpha/2+{1\over 2}} 
\over \big({z\over 2} + {\alpha\over 2}+{1\over 
2}\big)_{\alpha/2+{1\over 2}}} -2^\alpha. 
 \label{(A.9)} \end{equation}

 We now turn to the sum in the second factor of (\ref{(A.1)}).
  With some work one obtains
\begin{eqnarray}
 &&\sum^{\alpha-1}_{\scriptstyle j=0 
 \atop \scriptstyle j\text{ even}} (-1)^{\alpha-j} {(z+j+1)^{-
1}\over j!\,(\alpha-1-j)!} {\Gamma\big({1\over 2}(j+1)\big) \over 
\Gamma\big({1\over 2}(j+1)-\alpha\big)} 
 \nonumber\\*
&=& {(-1)^\alpha \big({1\over 2}-\alpha\big)_\alpha \over (z+1) 
(\alpha-1)!} \,{}_3F_2 \left(\matrix{{z\over 2} + {1\over 2}, -
{\alpha\over 2}+{1\over 2}, -{\alpha\over 2}+1\hfill\cr &;1\cr 
{z\over 2}+{3\over 2}, -\alpha+{1\over 2}\hfill\cr}\right). 
  \label{(A.10)} \end{eqnarray}

{\em Case 1: $\alpha$ is even.} 
 Again we apply (\ref{(A.2)}), setting 
 $a = -{\alpha\over 2} + {1\over 2}$, $b= {z\over 2} + {1\over 2}$ and 
 $-n = -{\alpha\over 2} +1$. 
 After a computation quite similar to the above we obtain:
\begin{equation}
 \sum^{\alpha-1}_{\scriptstyle j=0\atop
  \scriptstyle j\text{ even}} (-1)^{\alpha-j}
{(z+j+1)^{-1} \Gamma\big({1\over 2}(j+1)\big) \over j!\, (\alpha-1-
j)!\, \Gamma\big({1\over 2}(j+1)-\alpha\big)} 
 = 2^{-\alpha} {\big({z\over 2}+{\alpha\over 2}+1\big)_{\alpha/2} 
\over \big({z\over 2}+{1\over 2}\big)_{\alpha/2}} - 2^{-\alpha}. 
  \label{(A.11)} \end{equation}

{\em Case 2: $\alpha$ is odd.}
  Set $a = -{\alpha \over 2}+1$,
$b={z\over 2}+{1\over 2}$ and $-n = -{\alpha\over 2}+{1\over 2}$ 
 in (\ref{(A.2)}).
After another computation very similar to the previous ones we find 
 \begin{equation}
 \sum^{\alpha-1}_{\scriptstyle j=0\atop \scriptstyle j\text{ even}} 
 (-1)^{\alpha-j} {(z+j+1)^{-1} \Gamma\big({1\over 2}(j+1)\big) \over 
j!\,(\alpha-1-j)!\, \Gamma\big({1\over 2}(j+1)-\alpha \big)} 
 = 2^{-\alpha} {\big({z\over 2} +{\alpha\over 2}+{1\over 2}\big) 
_{\alpha/2 +{1\over 2}} \over \big({z\over 2} + {1\over 
2}\big)_{\alpha/2+{1\over 2}}} - 2^{-\alpha}. 
  \label{(A.12)} \end{equation}

Substituting (\ref{(A.8)})--(\ref{(A.12)}) back into (\ref{(A.1)}) 
yields the desired identity.

 \end{document}